\documentclass{aa}  

\usepackage{graphicx}
\usepackage{txfonts}

\usepackage[breaklinks,colorlinks,citecolor=blue]{hyperref}

\usepackage{ulem}
\usepackage{array}
\usepackage{lmodern}

\bibpunct{(}{)}{;}{a}{}{,} 

\newif\iffigs
\newif\iffigscl
\newif\iffigstest
\newif\iflabs

\figstrue

\figscltrue

\figstestfalse

\labsfalse

\begin{document} 


\newif\iffigs

\figstrue

\newcommand{\itii}[1]{{#1}}
\newcommand{\franta}[1]{\textbf{#1}}
\newcommand{\frantaii}[1]{\textbf{\color{green} #1}}
\newcommand{\itiitext}[1]{{#1}}

\newcommand{\eq}[1]{Eq. (\ref{#1})}
\newcommand{\eqp}[1]{(Eq. \ref{#1})}
\newcommand{\eqq}[1]{Eq. \ref{#1}}
\newcommand{\eqb}[2]{eq. (\ref{#1}) and eq. (\ref{#2})}
\newcommand{\eqc}[3]{eq. (\ref{#1}), eq. (\ref{#2}) and eq. (\ref{#3})}
\newcommand{\refs}[1]{Sect. \ref{#1}}
\newcommand{\reff}[1]{Fig. \ref{#1}}
\newcommand{\reft}[1]{Table \ref{#1}}

\newcommand{\datum} [1] { \noindent \\#1: \\}
\newcommand{\pol}[1]{\vspace{2mm} \noindent \\ \textbf{#1} \\}
\newcommand{\code}[1]{\texttt{#1}}
\newcommand{\figpan}[1]{{\sc {#1}}}

\newcommand{\nbdvi}{\textsc{nbody6} }
\newcommand{\nbdvid}{\textsc{nbody6}}
\newcommand{\flash}{\textsc{flash} }
\newcommand{\flashd}{\textsc{flash}}
\newcommand{\sfe}{\mathrm{SFE}}
\newcommand{\mum}{$\; \mu \mathrm{m} \;$}
\newcommand{\rop}{$\rho$ Oph }
\newcommand{\HT}{$\mathrm{H}_2$}
\newcommand{\Halpha}{$\mathrm{H}\alpha \;$}
\newcommand{\HI}{H {\sc i} }
\newcommand{\HII}{H {\sc ii} }
\renewcommand{\deg}{$^\circ$}

\newcommand{\dd}{\mathrm{d}}
\newcommand{\acosh}{\mathrm{acosh}}
\newcommand{\sign}{\mathrm{sign}}
\newcommand{\cex}{\mathbf{e}_{x}}
\newcommand{\cey}{\mathbf{e}_{y}}
\newcommand{\cez}{\mathbf{e}_{z}}
\newcommand{\cer}{\mathbf{e}_{r}}
\newcommand{\ceR}{\mathbf{e}_{R}}

\newcommand{\llg}[1]{$\log_{10}$(#1)}
\newcommand{\pder}[2]{\frac{\partial #1}{\partial #2}}
\newcommand{\pderrow}[2]{\partial #1/\partial #2}
\newcommand{\nder}[2]{\frac{\dd #1}{\dd #2}}
\newcommand{\nderrow}[2]{{\dd #1}/{\dd #2}}

\newcommand{\Cmiii}{\, \mathrm{cm}^{-3}}
\newcommand{\Gcmii}{\, \mathrm{g} \, \, \mathrm{cm}^{-2}}
\newcommand{\Gcmiii}{\, \mathrm{g} \, \, \mathrm{cm}^{-3}}
\newcommand{\Kms}{\, \mathrm{km} \, \, \mathrm{s}^{-1}}
\newcommand{\Si}{\, \mathrm{s}^{-1}}
\newcommand{\Esi}{\, \mathrm{erg} \, \, \mathrm{s}^{-1}}
\newcommand{\Ee}{\, \mathrm{erg}}
\newcommand{\Yr}{\, \mathrm{yr}}
\newcommand{\Kyr}{\, \mathrm{kyr}}
\newcommand{\Myr}{\, \mathrm{Myr}}
\newcommand{\Gyr}{\, \mathrm{Gyr}}
\newcommand{\Msun}{\, \mathrm{M}_{\odot}}
\newcommand{\Rsun}{\, \mathrm{R}_{\odot}}
\newcommand{\Pc}{\, \mathrm{pc}}
\newcommand{\Kpc}{\, \mathrm{kpc}}
\newcommand{\Mpc}{\, \mathrm{Mpc}}
\newcommand{\Sd}{\Msun \, \Pc^{-2}}
\newcommand{\Ev}{\, \mathrm{eV}}
\newcommand{\Kk}{\, \mathrm{K}}
\newcommand{\Au}{\, \mathrm{AU}}
\newcommand{\Mas}{\, \mu \mathrm{as}}

\newcommand{\logP}[1]{$\rm{log}_{10}(P[\rm{days}]) #1$}
\newcommand{\pcci}{p_{\rm CC} (M_{\rm ecl})}
\newcommand{\pccii}{p_{\rm CC} (M_{\rm ecl}, t)}
\newcommand{\pcciii}{p_{\rm CC} (m_{\rm Ceph})}

   \title{On the dynamical evolution of Cepheid multiplicity in star clusters and its implications for B-star multiplicity at birth}

   \authorrunning{Dinnbier, Anderson, and Kroupa}
   \titlerunning{Binary and multiple Cepheids in cluster environment}

   \author{Franti\v{s}ek Dinnbier\inst{1},
          Richard I. Anderson\inst{2}
          \and
          Pavel Kroupa\inst{1,3}
          }

   \institute{
             Astronomical Institute, Faculty of Mathematics and Physics, Charles University in Prague, V Hole\v{s}ovi\v{c}k\'{a}ch 2, 180 00 Praha 8, Czech Republic \\
             \email{dinnbier@sirrah.troja.mff.cuni.cz} \\
             \email{pavel.kroupa@mff.cuni.cz}
         \and
              Institute of Physics, \'Ecole Polytechnique F\'ed\'erale de Lausanne (EPFL), Observatoire de Sauverny, 1290 Versoix, Switzerland \\
              \email{richard.anderson@epfl.ch}
         \and
             Helmholtz-Institut f\"{u}r Strahlen- und Kernphysik, University of Bonn, Nussallee 14-16, 53115 Bonn, Germany
             }

   \date{Received \today; accepted ??}

\abstract
  {Classical Cepheid variable stars provide a unique probe of binary evolution in intermediate-mass stars over the course of several tens to hundreds of Myr. 
   In addition, understanding binary evolution with the inclusion of cluster dynamics is desirable 
   for obtaining a more complete picture of these stars, especially as they play a vital role in distance determinations.
   }
   {We studied the binary and multiple properties of Cepheids, assuming that all mid-B stars form in binaries inside star clusters. 
    We also estimated the birth multiplicity of mid-B stars by comparing the observed multiplicity statistics of Cepheids 
    with models based on particular assumptions.
  }
  {The clusters were modelled with the  \nbdvi code, including synthetic stellar and binary evolutionary tracks. 
   The Cepheids were identified from their position on the Hertzsprung-Russell diagram. 
  }
  {
   The dynamical cluster environment results in a higher binary fraction among the Cepheids that remain in star clusters ($\approx 60$\%) than 
   among the Cepheids which have escaped to the field ($\approx 35$\%). 
   The fraction of Cepheids in triples ($\approx 30$\% and $\approx 10$\% in clusters and field, respectively) follows the same trend. 
   In clusters, the binary, triple, and multiple fraction decreases with increasing cluster mass.
   More massive clusters have binaries of shorter orbital periods than lower mass clusters and field Cepheids.
   Mergers are very common with $\approx 30$\% of mid-B stars not evolving to Cepheids because of the interaction with their companion.
   Approximately $40$ \% of Cepheids have merged with their companion, and the merger event impacts stellar evolution, so that $\approx 25$\% of 
   all Cepheids occur at an age by more than $40$\% different than what would be expected from their mass and the current cluster age; 
   the expected age of Cepheids can differ from the age of their host cluster. 
   Our models predict that one in five Cepheids is the result of a merger between stars with mass below the lower mass limit for Cepheids;
   in clusters, these objects occur substantially later than expected from their mass.  
   Approximately $10$\% of binary Cepheids have a different companion from the zero-age main sequence (ZAMS) one, and 
   $\approx 3$ to $5$\% of all Cepheids have a compact companion ($\approx 0.15$ \% of all Cepheids are accompanied by a black hole). 
  }
  {
   The binary fraction derived from our simulations (42\%) underestimates the observed binary Cepheid fraction by approximately a factor of 2. 
   This suggests that the true multiplicity fraction of B-stars at birth could be substantially larger than unity and, thus, 
   that mid-B stars may typically form in triple and higher order systems.
  }
 

   \keywords{open clusters and associations: general, Stars: variables: Cepheids, binaries: general
               }

   \maketitle
%

\section{Introduction}


Classical Cepheids (henceforth, Cepheids) are evolved intermediate-mass stars best known for their high-amplitude pulsations 
and application as standard candles, thanks to Leavitt's law \citep[period-luminosity relation]{Leavitt1912}.
The B-star progenitors of Cepheids feature a frequency of companions greater than unity and recent astrometric studies 
based on the ESA mission \textit{Gaia} \citep{Gaiamission,GaiaDR2,GaiaEDR3} imply a Cepheid binary fraction on the order of $80\%$ \citep{Kervella2019a,Kervella2019b}. 
By including other types of binaries (UV observations, X-ray observations), 
\citet{Evans2022} have provided a comprehensive lower estimate of the binary fraction to be $57 \pm 12$ \%. 
Qualitatively, it is clear that short-period binaries are influenced (mass transfer or common envelope evolution) by the secular evolution 
of the primary star towards the red giant phase, which precedes the core He-burning blue loop phase where the vast majority of Cepheids are observed.
However, observational selection effects related to the generally large distances of Cepheids (hundreds to thousands of pc), long orbital periods ($\gtrsim 1$\,yr), and high contrast ratios at optical and near infrared wavelengths are challenging with respect to devising a detailed characterization of the Cepheid companion population and its orbital characteristics.

Substantial efforts to observationally determine the binary fraction of Cepheids have been made, resulting in a large body of accessible literature 
on the subject \citep[cf. the binary Cepheids database][]{Szabados}. 
For the sake of concision, it is sufficient to mention only a few of the most relevant recent studies based on high-quality data:   
line-of-sight evidence for orbital motion comes primarily from radial velocity (RV) observations (cf. \citealt{Anderson2017RVs} for a recent review), 
and, in a couple of cases, from pulsation timing \citep{Csornyei2022}.
\citet{Evans2015} determined a magnitude-limited binary fraction for stars with orbital periods $\lesssim 20$\,yr of $29\pm8\%$ 
based on 40 Milky Way Cepheids brighter than a  magnitude of 9. 
The systematic search for spectroscopic binary Cepheids \citep{Shetye2024} based on data from the {\tt VELOCE} project \citep{Anderson2024} 
has reassessed the evidence for binary candidates and identified many new binaries. 
This has sharpened the spectroscopic binary fraction to $29.6 \pm 3.4\%$, with $15.2 \pm 2.4\%$ having orbital periods shorter than $10$\,yr. 
\citet{Shetye2024} further found the shortest-period binary Cepheid in the Milky Way, R Crucis, with a period of $237$\,d, 
provided 30 homogeneously determined orbital solutions, and identified 17 triple systems with distant, visual companions of unknown period.


Astrometry offers a completely complementary and very detailed view of Cepheid multiplicity. 
For example, \citet{Kervella2019a} used varying proper motion vectors across the epochs 
from \textsc{hipparcos} and \textit{Gaia} DR2 to infer the presence of unseen companions and determine a binary fraction of $\gtrsim 80\%$ for Cepheids, 
more consistent with the high multiplicity fraction of B-stars. 
In a companion paper, \citet{Kervella2019b} further identified likely wide (resolved) companions via proper motion pairs.

The search for resolved companions requires techniques of resolving very fine angular scales due to the generally large distances of Cepheids (hundreds of pc to kpc). 
From the ground, the most successful techniques employed include lucky imaging \citep{Gallenne2014} and long-baseline 
interferometry \citep[e.g.][]{Kervella2019b}, which have allowed detection of companions with $H$-band magnitude difference of up to $5.5$.
From space, HST's WFC3 camera has been used to identify close-in ($< 5"$ separation) companions of a magnitude-limited sample of 70 Cepheids and 
revealed that all the 7 Cepheids with resolved companions closer than $2"$ were also spectroscopic binaries, often with significant eccentricity \citep{Evans2020}. 
A companion survey of more distant companions ($> 5"$ separation) \citep{Evans2016} matched with XMM-Newton observations 
to identify young companion stars found that X-ray active (i.e. young) 
companions of the 14 Cepheids with available X-ray observations were all closer than $4000 \Au$. 
More distant resolved companions identified in the vicinity of $\delta$~Cephei 
(which is also a low-amplitude spectroscopic binary; cf. \citealt{Anderson2015}) and 
S~Normae \citep{Irwin1955} were explained as coeval stars originating in the same birth cluster or association.

The properties of companions to Cepheids and their multiplicity frequency reflect the history of their evolution 
since their birth as mid-B stars within clusters embedded in molecular clouds \citep{Lada2003,Zinnecker2007}, followed by 
cloud dispersal due to stellar feedback  and the dynamical evolution of the exposed cluster in the galactic tidal field. 
This includes stellar encounters with other stars inside the natal cluster and stellar evolution of the binary components, whose importance increases
when the stars evolve beyond the main sequence (MS). 
The influence of stellar evolution was estimated by \citet{Neilson2015a} by means of binary star population synthesis, who found 
that the presence of a companion leads to a lower limit in Cepheid orbital periods of approximately one year
and that the destruction of closer-in companions implies that the mid-B star binary fractions should be about $15-30\%$ larger than the measured Cepheid binary fraction. 
More recently, \citet{Karczmarek2022} investigated the companion properties of a population of Cepheids in environments of different metallicity.

However, the influence of the dynamical environment of the Cepheid birth cluster has not been investigated yet.
Based on theoretical estimates and general results of N-body modelling, it has been
firmly established that stellar encounters of binaries with other single and binary stars in the cluster can change the orbital period and eccentricity 
of the binary by hardening, softening and dissolution, according to Heggie's law \citep{Heggie1975,Hills1975,Heggie2003}.
These interactions also involve resonances, exchanges, or collisions \citep[e.g.][]{Saslaw1974,Hut1983,Sigurdsson1993,Fregeau2004,Ginat2020} 
and they often result in fast ejections of single or binary stars, which are observed after their escape from the cluster 
as walkaway \citep{deMink2014,Schoettler2019} or 
runaway stars \citep[e.g.][]{Leonard1988,Leonard1990,Hoogerwerf2001,Eldridge2011,Fujii2011,Maiz2018}. 
The ability of a cluster to eject stars increases with the cluster density \citep{Hut1983,Binney2008,Oh2015} 
and with the presence of primordial binaries \citep{Perets2012,Oh2015,Oh2016}.
Some of the binaries containing Cepheids are likely to capture another star as an outer companion, thereby forming a hierarchical triple.
The outer orbits in triple stars are able to influence the orbital parameters of the inner binary 
via Kozai-Lidov oscillations \citep{Kozai1962,Lidov1962},
bringing the binary components closer together at their pericentre, whereby the orbit is circularised by tidal forces.
This mechanism produces short-period, low-eccentricity inner binaries \citep{Kiseleva1998,Fabrycky2007}.

In Paper I of this series \citep{Dinnbier2022}, we investigated the occurrence of Cepheids in star clusters and the field using N-body simulations. 
Here, we analyse the same suite of simulations with respect to the properties of Cepheid multiplicity. 
Stellar evolution in \nbdvid, in particular with respect to binary evolution (e.g. tidal circularisation, common envelope phase), has been implemented as 
described by \citet{Hurley2000} and \citet{Hurley2002}, which also form the basis 
of stellar evolution calculations for the analysis due to \citet{Neilson2015a} and \citet{Karczmarek2022}. 
A key difference between these two works and our study is that our N-body simulations consider the full 
cluster dynamical picture for a large range of birth environments, spanning from large bound clusters
to field Cepheids mostly originating from dispersed low-mass clusters. 
For example, this also allows us to assess the occurrence frequency of triples and higher-order systems that form dynamically after the zero-age main sequence (ZAMS).

\section{Initial conditions for binary stars}

\label{sInitCond}

This work utilises the extensive library of simulations documented in Paper I, which also describes 
the initial conditions for star clusters and numerical methods used in the code \nbdvi \citep{Aarseth1999,Aarseth2003}. 
Here, we provide a detailed description only of the initial conditions for the binary population and 
its subsequent integration, as these features are key in the context of the present work. 

All stars are formed in binaries at the beginning of the simulations. 
This means that all triple and higher-order systems in our simulations are formed only by capture events. 
Single stars, which are later paired to binaries, are drawn from the initial mass function (IMF) of \citet{Kroupa2001a}. 
To reflect the observed differences in the distribution functions for the orbital period, eccentricity, and 
mass ratio between early-type and late-type stars \citep{Moe2017,Offner2022}, 
we used two types of the initial distributions for binaries according to their primary mass, $m_1$; 
below and above the primary threshold mass, $m_{\rm 1,thr} = 3 \Msun$. 
The choice of $m_{\rm 1,thr}$, which is slightly lower than the usual value of $\approx 5 \Msun$ \citep{Kroupa2013}, ensures that 
all stars that evolve into Cepheids (minimum mass of $3.2 \Msun$ for the considered metallicity range) have their orbital parameters and mass ratios 
drawn from the same distribution functions (i.e. those for early-type stars), and they avoid the sudden transition to the other set of distribution 
functions, which occurs at $m_{\rm 1,thr}$. 
The initial distributions for orbital periods for binaries with $m_1 < m_{\rm 1,thr}$ and $m_1 > m_{\rm 1,thr}$ are 
shown in \reff{fCephBin} by the solid black and magenta lines, respectively.

Binaries with the primary of mass $m_1 < m_{\rm 1,thr}$ are generated with the birth distribution of orbital periods and 
eccentricities according to \citet{Kroupa1995a}. 
The interaction of the two protostars with each other and with the protostellar material that remains in the system (pre-MS eigenevolution) 
is approximated by the method of \citet{Kroupa1995b} with the update attributed to \citet{Belloni2017}. 
Pre-MS eigenevolution takes place on the time-scale of $10^5$ years, 
transforming the birth binary population to the initial binary population.
Pre-main-sequence (pre-MS) eigenevolution is applied on each binary individually before the start of the simulations;
the initial binary population is then taken as the initial conditions for the integration in \nbdvid, 
which starts at time $t_{\rm 0}$.
For simplicity, it is assumed that all stars are already at their ZAMS at a time,  $t_{\rm 0}$;
whereas, in nature, it takes tens of Myr for solar type stars to reach the ZAMS (for the dynamical consequences of this approximation, we refer to \citealt{Railton2014}).

The birth distribution of binaries is a theoretical concept (analogically as the IMF), 
which is not likely to be observed directly because of the large extinction and rapid evolution of these systems. 
However, the adopted form of the birth distribution, after being processed via pre-MS eigenevolution and then cluster dynamics (stimulated evolution), 
closely reproduces the observed properties (binarity, orbital period distribution) 
of stars in clusters and in the field \citep[e.g.][]{Kroupa1995b,Kroupa2001b,Marks2012,Belloni2018}, 
which is accessible to observations. 

Binaries with a primary of mass $m_1 > m_{\rm 1,thr}$ were generated with the birth distribution of the mass ratio, orbital periods, and 
eccentricity taken from  \citet{Belloni2017}, which differs from that of their lower mass counterparts. 
These distributions are based mainly on the observations due to \citet{Sana2012,Kobulnicky2014,Sana2014,Aldoretta2015}.
In particular, the orbital period distribution function, $f_{\rm p}$, for massive stars is given by \citep{Sana2012}:
\begin{equation}
f_{\rm p} (\log_{10}P) = 0.23 \; (\log_{10}P)^{-0.55}, 
\label{eSana}
\end{equation}
where orbital periods occupy the range $\log_{10}P \in (0.15,6.7)$, where $P$ is in days.
In contrast to the lower mass binaries, the more massive binaries have no pre-MS eigenevolution, namely, their birth distribution is 
the same as the initial distribution. 
This approximation is motivated by the much faster pre-MS evolution of massive stars \citep[e.g.][]{Stahler2004},  
uncertainties in massive star formation \citep[e.g.][]{Zinnecker2007,Peters2010b}, and the fact that a study addressing the influence of cluster environment 
on properties of massive binaries (an analogy to the iterative solution of \citealt{Kroupa1995a} for late type binaries) has not been undertaken so far. 
Instead, we assume that massive binaries form with the same properties as the observed ones. 

When initialising the binary population, it is of paramount importance (as stressed in \citealt{Kroupa2018b}) to not affect the IMF. 
This is achieved by first sampling stars from the stellar IMF and only then distributing the already sampled stars into the binaries. 
We note that initializing the binary population by first sampling only the primaries from the IMF and then generating the secondary mass 
according to an assumed mass-ratio distribution (as is often done) is likely to affect the shape of the IMF, which can lead to contradictions with the star counts.

\iffigscl
\begin{figure}
\includegraphics[width=\columnwidth]{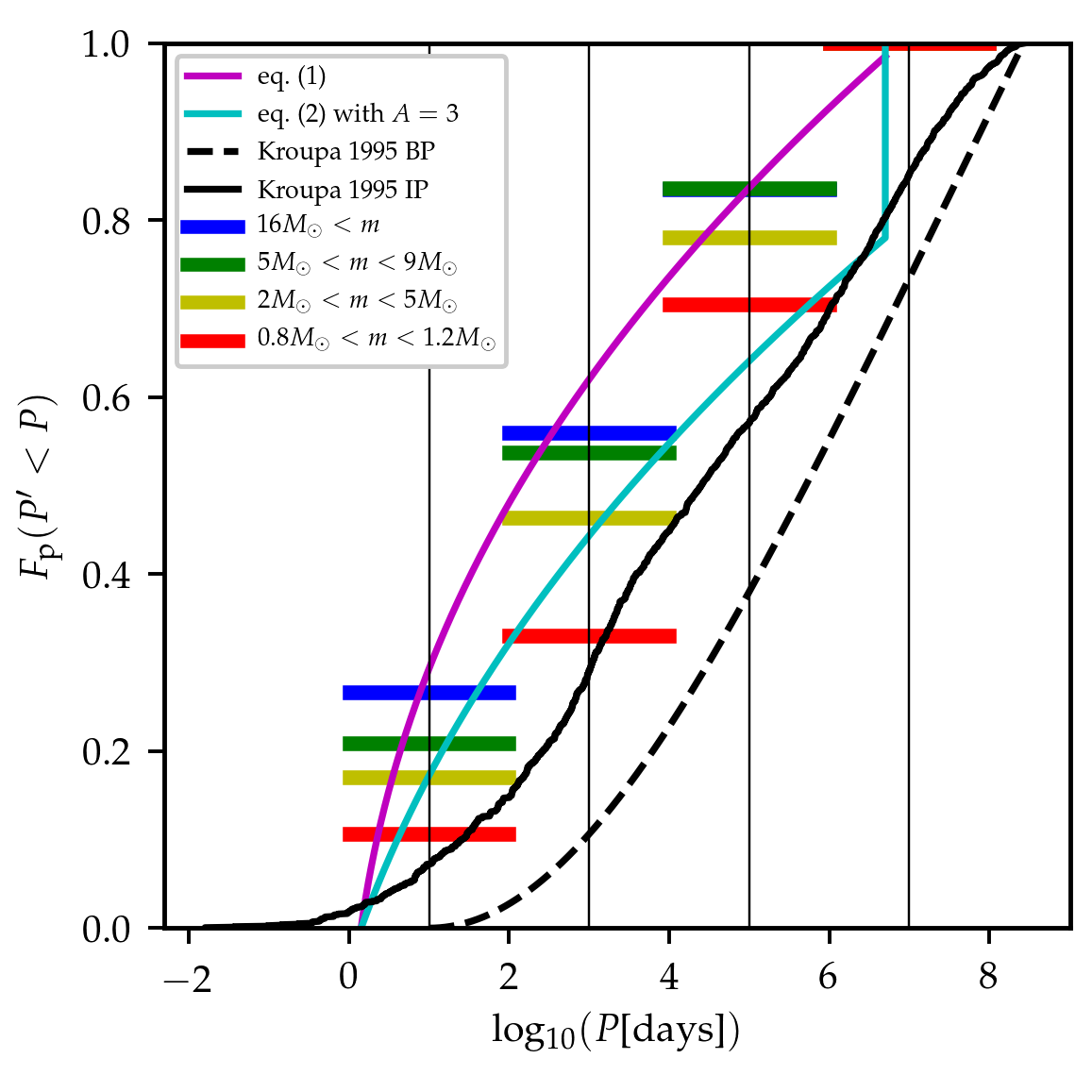}
\caption{Cumulative distribution functions of orbital periods of binary stars according to the primary mass.
The observed distributions for primaries in specific mass intervals are shown by bars \citep{Moe2017}: 
O-type stars ($m > 16 \Msun$; blue bars), mid B-type stars ($5 \Msun < m < 9 \Msun$; green), late B- and A-type stars ($2 \Msun < m < 5 \Msun$; yellow), 
and solar type stars ($0.8 \Msun < m < 1.2 \Msun$; red).
The theoretical birth distribution of \citet{Kroupa1995a} for late type stars ($m < 3 \Msun$; dashed black line) is transformed by pre-MS eigenevolution to the initial distribution (solid black line). 
The initial distribution for O-stars according to \eq{eSana} \citep{Sana2012} and the wider distribution of \eq{eWide} with $A = 3$, 
which is the widest estimate on the Cepheid orbital periods, is plotted by the magenta and cyan line, respectively.
}
\label{fCephBin}
\end{figure} \else \fi


The orbital parameters of binaries change not only due to the stellar evolution in binary stars, 
but also due to dynamical encounters with other stars and binaries within the cluster. 
According to the Heggie's law \citep{Heggie1975,Binney2008}, sufficiently close binaries harden (reduce their orbital period $P$) due to dynamical encounters, 
bringing their companion closer so that the binary merges in some cases during post-MS 
evolution, although it would not have merged had it evolved in the absence of the other stars in the same cluster. 
A sufficiently large mass increase during such a merger might increase the mass of the star so much that it does not become a Cepheid. 
In contrast, wider binaries can soften with the opposite consequence of being dissociated. 
These examples point out that the fraction of stars which actually become Cepheids
is given not only by the distribution function of their initial orbital periods, but also by the internal stellar dynamics of their host clusters. 

Here, the binaries are investigated in star clusters orbiting the galaxy on circular trajectories at galactocentric radii $R_{\rm g} = 4 \Kpc$, 
$R_{\rm g} = 8 \Kpc$, and $R_{\rm g} = 12 \Kpc$ for a metallicity of $Z = 0.014$. 
Positioning the cluster orbits at different radii results in a different degree of confinement due to 
the gravitational potential of the galaxy (i.e. different tidal radius of the cluster), 
which is represented by the Milky-Way like gravitational potential \citep{Allen1991}.
At each galactocentric radius, we have studied clusters across a wide mass spectrum: from $M_{\rm ecl} = 50 \Msun$ to $6400 \Msun$ ($M_{\rm ecl}$ is the total stellar mass of the initially gas embedded cluster)
in steps increasing $M_{\rm ecl}$ by a factor of $2$. 
Less massive clusters are realised more often than the more massive ones to obtain more robust statistics for Cepheids; clusters of mass 
$6400 \times 2^{-i} \Msun$ are realised $2^{i+1}$ times ($i$ being an integer from $0$ to $7$) with a different random seed for initial conditions. 
We also studied the possible influence of metallicity by calculating additional sets of models at $R_{\rm g} = 8 \Kpc$ with $Z = 0.002$ and $Z = 0.006$. 
These models, which all include internal cluster dynamics, are referred to as "standard models."
We refer to Paper I for more details. 

To separate the influence of stellar evolution and cluster dynamics on binaries, 
we performed additional simulations with cluster radii large enough (half mass radius $\approx 100 \Pc$) for stellar dynamical processing 
of the binary population to be negligible, but with a stellar evolution identical to that in the standard models.
These models, which we call "control models", are calculated for the 
same metallicity as the standard models ($Z = 0.014$, $Z = 0.006,$ and $Z = 0.002$). 
Control models consist of the same cluster masses $M_{\rm ecl}$ as standard models (from $50 \Msun$ to $6400 \Msun$ in steps increasing $M_{\rm ecl}$ 
by a factor of $2$) and with the same number of realisations (e.g. we calculated two times as many clusters of initial mass $M_{\rm ecl}$ 
than clusters of initial mass $2M_{\rm ecl}$). 
The reason of using the same cluster mass in control models is to generate the initial binary population of identical properties as in the standard models; 
the least massive clusters ($M_{\rm ecl} \lesssim 200 \Msun$) do not form the most massive Cepheids because we assume that 
the mass of the most massive star is a function of the cluster mass (\citealt{Weidner2010,Yan2023}; 
see also Table 1 of Paper I for maximum stellar mass, $m_{\rm max}$, forming in a cluster of mass $M_{\rm ecl}$). 
Control simulations follow the evolution not only for Cepheids that evolve from stars of the appropriate initial mass range $(m_{\rm min,Ceph}, m_{\rm max,Ceph})$, 
but also for Cepheids which originate from a star below this mass range, which gains mass by merging with its companion. 
For this purpose, the control runs are calculated up to $2 \Gyr$, which is long enough for the least massive stars which can evolve to 
Cepheids in a mass conserving merger of two $m_{\rm min,Ceph}/2 \approx 1.6 \Msun$ stars. 


The typical mass of stars that evolve to Cepheids (mainly mid-B stars) is lower than the mass of O stars, whose 
binary period distribution was found by \citet{Sana2012} (c.f. \eqq{eSana}). 
Generally, observations show that the median of the orbital period distribution of early-type stars increases 
with decreasing primary mass (\citealt{Duchene2013,Moe2017,Offner2022}; \reff{fCephBin}), 
so the stars that evolve into Cepheids have a wider orbital period distribution than O-stars; 
the assumption of the same orbital period distribution 
for all stars with $m > 3 \Msun$ according to \eq{eSana} is a simplification. 
To access the influence of this approximation on the binary evolution of stars with $3 \Msun \lesssim m \lesssim 12 \Msun$, 
we perform an additional control run of binaries with a wider orbital period distribution%
\footnote{This distribution is the original Sana's distribution \eqp{eSana} with semi-major axes inflated by a factor of $A$. 
With $A = 1$, \eq{eWide} becomes \eq{eSana}.}
,
\begin{equation}
f_{\rm p} (\log_{10}P) = 0.23 \; \left( \log_{10}P + \frac{3}{2}\log_{10}(A) \right)^{-0.55},  
\label{eWide}
\end{equation}
with the parameter $A = 3$.
The purpose of this distribution is to approximate the 
orbital period distribution for stars in the $(2 \Msun, 5 \Msun)$ range (cyan line in \reff{fCephBin}). 
According to the modified Sana's distribution with $A = 3$, $50$\% of binaries have $\log_{10} (P$[days]$) < 3.5$, 
while according to the original Sana distribution \eqp{eSana}, 
$50$\% of binaries have $\log_{10} (P$[days]$) < 2.1$. 
Models with binary period distribution of \eq{eWide} are investigated in \refs{ssWidePer}.

During the simulation, \nbdvi regularises compact binaries, and integrates them according to the Kustaanheimo-Stiefel method \citep{Kustaanheimo1965}. 
Hierarchical systems, which form dynamically in the course of the simulation, 
are treated either by the three-body regularization method of \citet{Aarseth1974a}, if the inner binary is perturbed by the outer body, 
or are treated as single objects (\citealt{Mardling1999}) if the inner binary is sufficiently isolated from the outer body, 
which considerably speeds up the calculation.
\nbdvi takes into account single star \citep{Tout1996,Hurley2000} as well as binary star evolution \citep{Aarseth1996a,Hurley2002}, which is 
approximated by synthetic stellar evolutionary tracks. 
Collisions and subsequent merges between stars are evaluated according to the criterion of \citet{Kochanek1992}. 
The code enables modelling of the common envelope evolution and Roche-lobe mass transfer, 
where the Roche radius is estimated by the formula of \citet{Eggleton1983}. 
The compact remnant after a supernova explosion receives a kick in a randomly selected direction and of the velocity drawn from a Maxwell 
distribution with 1D dispersion $\sigma = 40 \Kms$ \citep{Aarseth1996b,Hansen1997}.
The kick might result in binary disruption.

\section{Evolutionary channels of Cepheids in binaries}

\label{sChannels}

For the sake of clarity, we introduce terminology for Cepheids originating from stars of different initial masses or with a different history of 
interaction with their companions. 
Several terms of the present terminology differ from the terminology adopted in Paper I; 
to avoid confusion, we present a comparison of the terminologies in \reft{tTerminology}.
We refer to the stars formed within the mass range from which Cepheids evolve in isolation 
as prospective Cepheids (hereafter, ProCeps). 
The ZAMS mass of ProCeps slightly increases with metallicity 
from the range $(m_{\rm min,Ceph}, m_{\rm max,Ceph}) \in (3.2, 11) \Msun$ for $Z = 0.002$
to $(m_{\rm min,Ceph}, m_{\rm max,Ceph}) \in (4.7, 13) \Msun$ for $Z = 0.014$ \citep{Anderson2016a}.
Given that a significant fraction of ProCeps are in close binaries \eqp{eSana}, these stars 
interact with their companions as they evolve out of the MS. 
Close binaries (\llg{$P $[days]} $\lesssim 2.6$; \citealt{Neilson2015a,Moe2017}) merge or undergo Roche-lobe mass transfer 
or common envelope evolution, which in some cases changes the 
evolution of the final object to such an extent that it does not become a Cepheid or it becomes a Cepheid at a significantly 
different time than expected from its mass at the ZAMS.

The considered mass changes are sketched in \reff{ftimeCartoonCeph}, where 
ProCeps are located between the blue-dashed horizontal lines (mass is on the ordinate) at their ZAMS. 
Some ProCeps evolve to the Cepheid stage, which is shown by the hatched area. 
Some of the ProCeps that evolve to Cepheids do not interact with their companions, they effectively evolve in isolation (CanCeps; violet line), 
while others experience mass gain (InterCeps-I; grey line) or mass loss (InterCeps-II; grey line) to their companions. 
The mass gain or mass loss can be due to merger or mass transfer with their companion. 
The Cepheids that evolve from ProCeps (i.e. CanCeps and InterCeps of both types) are refereed to as Cepheids originating from the expected mass range:\ RanCeps.
The more massive Cepheids have the tips of their blue loops outside the instability strip (IS).

Other ProCeps experience more significant interactions with their companions (e.g. merger, Roche-lobe mass transfer or common envelope evolution) 
that prevent them from reaching the Cepheid stage (ExCeps). 
If the ExCep gains mass during the interaction, it is defined to be of type I (brown line); if the ExCep loses mass during the interaction, 
it is defined to be of type II (orange line). 
Analogously, some stars with the ZAMS mass outside the interval $(m_{\rm min,Ceph}, m_{\rm max,Ceph})$ 
become Cepheids (AdCeps) either due to mass gain (type I; dark green) or mass loss (type II; light green) with respect to their companion. 
The method of automatic identification of Cepheids in \nbdvi simulations is described in sect. 3 of Paper I.

Before analysing the process of binary evolution leading to Cepheids, it is instructive to estimate the percentage of ProCeps that form 
as primary or secondary in a representative sample of a coeval stellar population.
Assuming the IMF of \citet{Kroupa2001a}, the mass ratio distribution between the primary and secondary of \citet{Belloni2017} 
and a star-burst producing embedded clusters distributed according to the mass function of \eq{eCSP} 
with the cluster mass range spanning from $50 \Msun$ to $6400 \Msun$, 
approximately 10 \% of ProCeps are secondaries where the primary is more massive than $m_{\rm max,Ceph}$,
50 \% of ProCeps are in a binary with another ProCep, and 
40 \% of ProCeps are primaries where the secondary is less massive than $m_{\rm min,Ceph}$. 



\iffigscl
\begin{figure}
\includegraphics[width=\columnwidth]{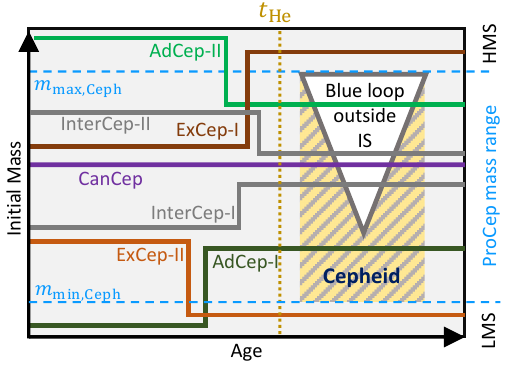}
\caption{
Schematic illustration of the considered mass changes due to binary evolution. 
The abscissa represents time; the ordinate represents the initial stellar mass, $m$. 
ProCeps, which have an initial mass range of $(m_{\rm min,Ceph}, m_{\rm max,Ceph})$ evolve towards Cepheids 
(without mass exchange:\ violet line or with mass exchange:\ grey lines), 
the state of which is marked by the hatched area. 
The beginning of core helium burning is indicated by the yellow dotted line.
See \refs{sChannels} for full explanation of the evolutionary channels and the terminology used in this paper.
}
\label{ftimeCartoonCeph}
\end{figure} \else \fi

\section{Evolution of binary and multiple Cepheids in star clusters}

\label{sCircumBin}

In star clusters, binary stars interact with the other cluster members, which can change the
initial orbital parameters of a binary before it starts evolving out of the MS.
Depending on the nature of the interaction,
the binary can harden (a decrease of orbital period $P$), soften (an increase of $P$), or be ionised (disrupted). 
For an idealised star cluster of half-mass radius $r_{\rm h}$ and consisting of $N$ equal mass stars, 
it is possible to derive an approximate watershed for the binary semi-major axis $a_{\rm hard}$ \citep{Heggie2003}:  
\begin{equation}
a_{\rm hard} = r_{\rm h}/N,
\label{eHardThr}
\end{equation}
where binaries with the semi major axis smaller than 
$a_{\rm hard}$ harden and binaries with the semi major axis larger than $a_{\rm hard}$ soften.
For comparison, in the control sets of simulations, binaries are subjected to their stellar (binary)
evolution only, being isolated from the possible influence of other binaries or single stars. 

When the quantities of interest are evaluated for a large sample of stars formed in a population of star clusters of different masses, 
we assume that the embedded cluster mass function (ECMF) takes the form 
\begin{equation}
\nder{N_{\rm ecl}(M_{\rm ecl})}{M_{\rm ecl}} \propto M_{\rm ecl}^{-\beta},
\label{eCSP}
\end{equation}
where $M_{\rm ecl}$ is the initial stellar mass of the embedded cluster and $\dd N_{\rm ecl}$ is the number of clusters in the cluster mass bin of size $\dd M_{\rm ecl}$. 
We adopt the slope $\beta = 2$ \citep{Lada2003,Bik2003,FuenteMarcos2004,Gieles2006}.
The mass of the embedded clusters spans the range of $(50 \Msun, 6400 \Msun)$. 
The typical number of Cepheids formed in the population of clusters in one set of models is $\approx 1700$ 
(for $Z = 0.014$; the number increases for models with sub-solar metallicities).
The ECMF mass slope and the range of cluster masses are the same as adopted in Paper I.


\begin{table*}
\caption{
Outcomes of binary evolution for ProCeps and Cepheids.
} 
\label{ttestRun}
\begin{tabular}{ccccccccccccccc}
$Z$ & $R_{\rm g} $ & CD & PD & $p_{\rm EC}$ & $p_{\rm soft}$ & $p_{\rm hard}$ & $p_{\rm C,coll,in}$ & $p_{\rm C,coll,out}$ & $p_{\rm mma}$ & 
$p_{\rm ACI}$ & $p_{\rm BC, in}$ & $p_{\rm BC, out}$ & $p_{\rm chg}$ & $p_{\rm cmp}$ \\
\hline
0.014 & 4 & Yes & \eq{eSana} &  0.30 &  0.19 &  0.07 &  0.48 & 0.36 & 0.25 & -   & 0.62 & 0.39 & 0.09 & 0.03  \\
0.014 & 8 & Yes & \eq{eSana} &  0.30 &  0.18 &  0.05 &  0.45 & 0.36 & 0.27 & -   & 0.62 & 0.34 & 0.08 & 0.03  \\
0.014 & 12 & Yes & \eq{eSana} &  0.29 &  0.18 &  0.06 &  0.44 & 0.38 & 0.29 & -  & 0.60 & 0.36 & 0.11 & 0.04  \\
\hline
0.006 & 8 & Yes & \eq{eSana} & 0.26 &  0.16 &  0.05 &  0.43 & 0.39 & 0.15 & -    & 0.70 &  0.46 & 0.08 & 0.05 \\
0.002 & 8 & Yes & \eq{eSana} & 0.27 &  0.15 &  0.07 &  0.54 & 0.46 & 0.16 & -    & 0.62 &  0.49 & 0.09 & 0.06 \\
\hline
0.014 & - & No & \eq{eSana} &  0.24 &  0.12 &  0.02 &  \multicolumn{2}{c}{0.40} & 0.20 & 0.17 & \multicolumn{2}{c}{0.50}  & -    & 0.05\\
0.006 & - & No & \eq{eSana} &  0.21 &  0.12 &  0.01 &  \multicolumn{2}{c}{0.41} & 0.15 & 0.11 & \multicolumn{2}{c}{0.56}  & -    & 0.08 \\
0.002 & - & No & \eq{eSana} &  0.23 &  0.11 &  0.03 &  \multicolumn{2}{c}{0.45} & 0.21 & 0.14 & \multicolumn{2}{c}{0.56}  & -    & 0.10 \\
\hline
0.014 & - & No & \eq{eWide} &  0.17 &  0.16 &  0.02 & \multicolumn{2}{c}{0.27} & 0.14 & -    & \multicolumn{2}{c}{0.59}   & -    & 0.07
\end{tabular}
\tablefoot{
The outcomes are listed as a function of metallicity $Z$, 
galactocentric radius of the cluster orbit $R_{\rm g}$, cluster dynamics (CD; Yes or No) and the (orbital) period distribution (PD). 
The values are calculated for the whole population of star clusters of the ECMF of \eq{eCSP}, 
which spans the initial cluster mass range $(50 \Msun, 6400 \Msun)$.
The properties of interest are as follows: $p_{\rm EC}$ - fraction of ExCeps (types I and II combined); 
$p_{\rm soft}$ - fraction of ProCeps which have softened between time $t_{\rm 0}$ and TAMS; 
$p_{\rm hard}$ - fraction of ProCeps which have hardened between time $t_{\rm 0}$ and TAMS; 
$p_{\rm C,coll,in}$, $p_{\rm C,coll,out}$ - fraction of Cepheids (in clusters and in the field respectively) which have coalesced with another star; 
$p_{\rm mma}$ (mismatched age) - fraction of Cepheids whose age, as calculated from their mass, differs by more than $40$\% from 
the age of the cluster (results from binary evolution); 
$p_{\rm ACI}$ - fraction of AdCeps type I (relative to the total number of ProCeps at time $0$);
$p_{\rm BC, in}$, $p_{\rm BC, out}$ - fraction of Cepheids (in clusters and in the field respectively) which are in a binary with $P < 10^{10}$ days; 
$p_{\rm chg}$ - fraction of binary Cepheids which have a different companion than at time $t_{\rm 0}$;
$p_{\rm cmp}$ - fraction of Cepheids with a compact companion.
}
\end{table*}

\subsection{Cepheids in binaries}

\label{ssBin}

\iffigscl
\begin{figure*}
\includegraphics[width=\textwidth]{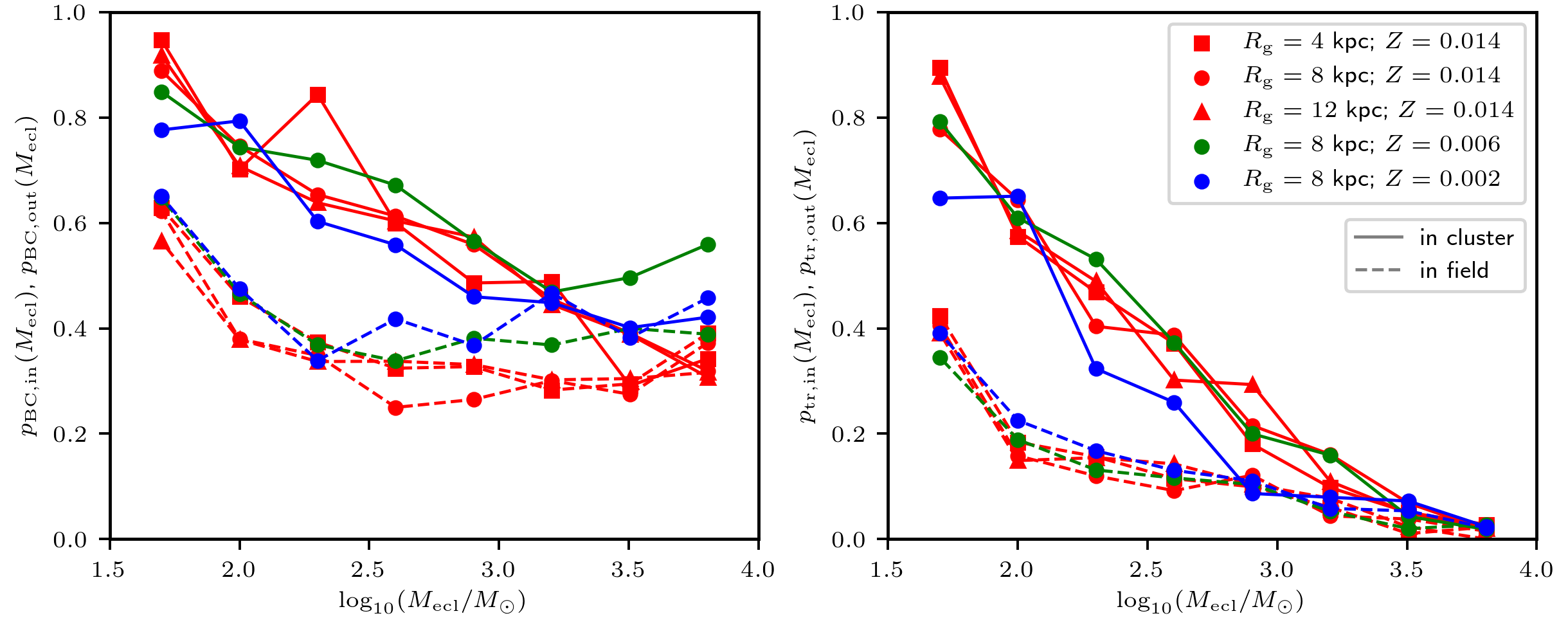}
\caption{
Fraction of Cepheids in binaries and triples as a function of the initial stellar mass of the embedded cluster where the Cepheids formed.
\figpan{Left panel:} Fraction of Cepheids in binaries of orbital period shorter than $10^{10}$ days.
The fraction for Cepheids occurring in clusters and in the field is plotted separately by solid and dashed lines, respectively. 
Clusters at different galactocentric radii are shown by symbols, and clusters of different metallicity are shown by colours. 
Some of the Cepheids in binaries are orbited by an outer body, namely, they are the inner binary of a triple. 
\figpan{Right panel:} Fraction of Cepheids in triples where the orbital period of the outer body is shorter than $10^{10}$ days. 
The meaning of the symbols is the same as in the left panel. 
}
\label{fBinHiaCep}
\end{figure*} \else \fi

\iffigscl
\begin{figure*}
\includegraphics[width=\textwidth]{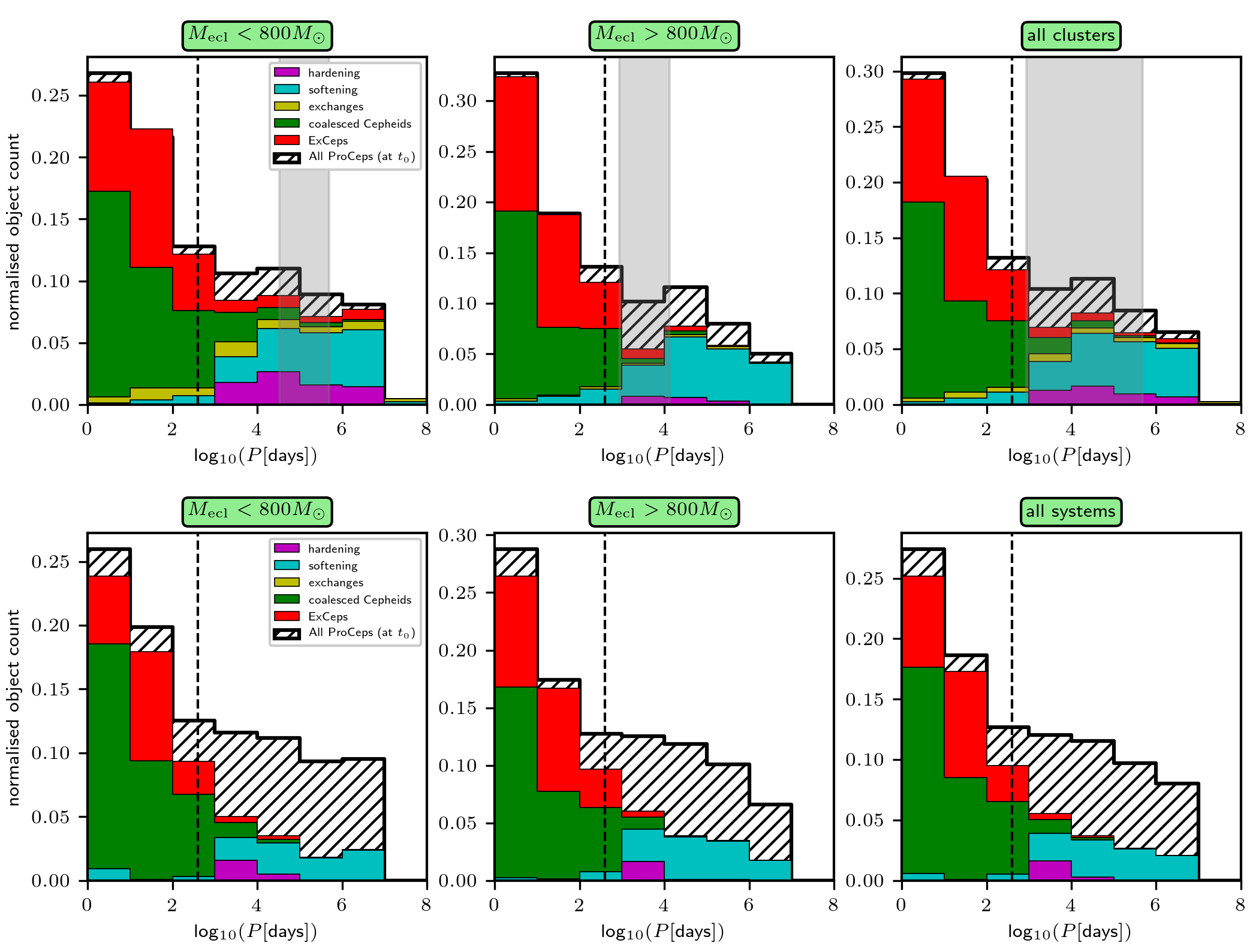}
\caption{
Outcome of binary evolution in ProCeps as a function of their initial orbital period. 
\figpan{Upper row:} The panels from left to right show the future of ProCeps formed in a population of star clusters 
of the assumed ECMF of \eq{eCSP} for clusters of mass $M_{\rm ecl} < 800 \Msun$, clusters with $M_{\rm ecl} \gtrsim 800 \Msun$, and all clusters combined. 
BiCeps which harden, soften or exchange their companions are shown by the magenta, cyan and yellow regions, respectively. 
Cepheids which have coalesced with their companion are shown by green regions, and ExCeps are shown by red regions. 
These evolutionary outcomes are shown as stacked histograms.
The initial orbital period distribution for all Cepheids is shown by the thick black histogram (the hatched regions represent the rest of BiCeps, which have not underwent 
any of the aforementioned evolutionary changes). 
We note that the colour-coded histograms are not additive; for instance an exchange that hardens at the same time 
is shown in both areas as if it were two separate binaries.
The orbital period of a binary can change between $t_{\rm 0}$ and the time when the star becomes a Cepheid (see e.g. \reff{fEccPlaneEvolvArr}).
The vertical black dashed lines (at $\log_{10} (P$[days]$) = 2.6$) show the upper orbital period for physical collision with their companion during the Cepheid phase.
The vertical grey bands represent the threshold orbital period $P_{\rm sep}$ between the soft and hard binaries for clusters of the particular mass range.
The plots are for a population of star clusters with $R_{\rm g} = 8 \Kpc$ and $Z = 0.014$. 
\figpan{Lower row:} Same as the upper row but for the control models with the same metallicity.
}
\label{fOrbPerChange}
\end{figure*} \else \fi

\iffigscl
\begin{figure}
\includegraphics[width=\columnwidth]{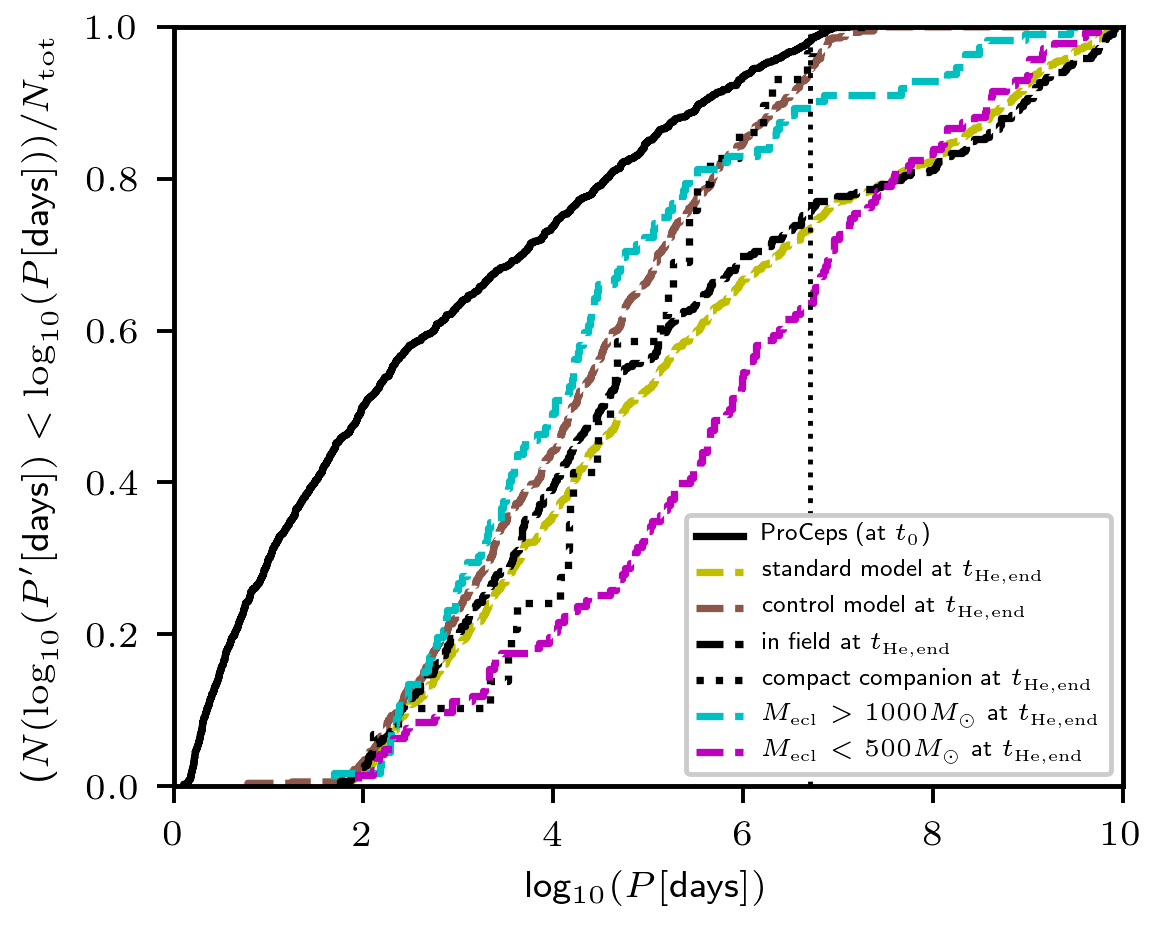}
\caption{
Cumulative distribution of orbital periods of ProCeps (taken at $t_{\rm 0}$; black solid line) and BiCeps (taken at $t_{\rm He,end}$; dashed lines). 
The orbital period distribution of BiCeps is shown separately for Cepheids 
in the lower (magenta dashed line) and more massive clusters (cyan dashed line), 
in the whole cluster population (yellow dashed line), 
in the field originating from the whole cluster population (black dashed line), 
and in the control model (brown dashed line). 
The orbital period distribution of Cepheids with a compact companion is indicated by the black dotted line.
The distributions are normalised to the total number of BiCeps in each group. 
There are almost no BiCeps with $\log_{10} (P \mathrm{[days]}) \lesssim 2$ 
because of coalescence with their companion before the onset of core helium burning. 
The vertical dotted line represents the upper limit on the initial orbital periods ($\log_{10} (P \mathrm{[days]}) \lesssim 6.7$).  
All BiCeps with an orbital period longer than that formed due to dynamical encounters in clusters.
}
\label{fperDist_v4}
\end{figure} \else \fi

We consider a Cepheid to be a member of a binary system (hereafter binary Cepheid; BiCep) if 
it is a component of a binary of an orbital period shorter than $10^{10}$ days (corresponding to a semi major axis of $1 \Pc$ and orbital 
period of $28 \Myr$ for an $8 \Msun$ binary)
both at the terminal-age main sequence (TAMS) and at the end of core helium burning ($t_{\rm He,end}$); the latter event occurs immediately after the Cepheid stage. 
This constitutes rather an upper limit on the binary orbital period because the orbital time is comparable to the time for a star to evolve to a Cepheid
and also the Jacobi radius of the binary ($\approx 3 \Pc$) in the Galactic potential is not much larger than its semi-major axis. 
Although some of the widest binaries are only temporarily bound, they do not increase the binary fraction significantly because 
most of the binaries have $\log_{\rm 10} (P$[days]$) \lesssim 8$ (see \reff{fperDist_v4} below for the orbital period distribution). 

\subsubsection{Fraction of binary Cepheids in clusters and in the field}

We adopt the definition of the fraction of binary Cepheids $p_{\rm BC}$ as the ratio of all Cepheids with at least one companion to the total number of Cepheids
\footnote{
Some of the binaries are thus orbited by one or more outer companions (i.e. triple and quadruple systems), 
but the two innermost stars would appear as a binary while different observing techniques (e.g. visual) would be needed for detecting the outer bodies. 
For this reason, we count them as binaries, and we study the outer companions separately in \refs{ssTriQ}.
}
.
We determine the fraction of BiCeps separately for Cepheids located in star clusters (fraction $p_{\rm BC, in}$) 
\footnote{As in Paper I, we classify a Cepheid as a cluster Cepheid if the star is closer than $10 \Pc$ from 
the density centre \citep{Casertano1985} of the cluster. 
If the Cepheid is further away, or the cluster is no longer gravitationally bound, we classify the Cepheid as a field Cepheid. 
\citet{CruzReyes2023} present the most up-to-date census of Cepheids in open clusters in the Milky Way.}
and in field ($p_{\rm BC,out}$).
For a population of stars originating from clusters of the ECMF of \eq{eCSP}, $p_{\rm BC, in} \approx 0.61$ 
is larger than $p_{\rm BC, out} \approx 0.36$ (\reft{ttestRun}). 
The binary fraction both in the clusters and in the field decreases with increasing cluster mass as shown in the left panel of \reff{fBinHiaCep}.
The influence of pure binary evolution is obtained from the control models; the coalescence with companions (in $44$\% of cases) and 
the disruption due to the supernova kick to primaries where Cepheids are secondaries (in $6.5$\% of cases), reduce the fraction 
of BiCeps to $p_{\rm BC} = 0.50$ (for $Z = 0.014$; \reft{ttestRun}).
Cluster dynamics slightly decreases the fraction of BiCeps, because the standard models have $p_{\rm BC} \approx 0.42$ (for $Z = 0.014$) in clusters and field combined.

\subsubsection{Changes in orbital period distribution}

\label{sssChangesPeriods}

\begin{table}
\caption{
Dependence of orbital periods of binary Cepheids on their environment or type of companion.
}
\label{tOrbPer}

\resizebox{\columnwidth}{!}{%
\begin{tabular}{lccc}
binary type & $Q_{\rm 50}$ & $Q_{\rm 75} - Q_{\rm 50}$ & $Q_{\rm 50} - Q_{\rm 25}$ \\
\hline
All ProCeps & 2.0 & 2.1 & 1.2 \\
All BiCeps & 4.8 & 2.1 & 1.4 \\
BiCeps in field & 4.5 & 2.1 & 1.2 \\
BiCeps in clusters & 5.3 & 2.0 & 1.8 \\
BiCeps in clusters of $M_{\rm ecl} < 500 \Msun$ & 5.9 & 1.4 & 1.3 \\
BiCeps in clusters of $M_{\rm ecl} > 1000 \Msun$ & 4.1 & 1.1 & 1.1 \\
BiCeps with compact companion & 4.6 & 0.9 & 0.5 \\
All outer triples including BiCeps & 8.5 & 0.8 & 1.4 \\
\hline
All BiCeps in control models & 4.2 & 1.2 & 1.1
\end{tabular}
}
\tablefoot{
$Q_{\rm 50}$ indicates the median of the orbital period (in $\log_{\rm 10}(P$[days]$)$).
The extent and asymmetry of the $\log_{\rm 10}(P)$ distribution is indicated by the difference between its median and its 
upper quartile $Q_{\rm 75}$ (75th percentile) and lower quartile $Q_{\rm 25}$ (25th percentile). 
The upper eight lines pertain to the standard models, 
while the last line pertains to the control models.
The results correspond to the population of clusters with $R_{\rm g} = 8 \Kpc$ and $Z = 0.014$.
}
\end{table}

In order to obtain the fraction of binaries which soften ($p_{\rm soft}$) or harden ($p_{\rm hard}$), 
we compare the orbital periods between time $t_{\rm 0}$ and the terminal age main-sequence $t_{\rm TAMS}$ 
of the binaries in which the ProCeps reside.
We choose the TAMS as the upper time limit because it is the time when the star starts increasing its radius substantially, which 
increases the probability of the interaction with its companion, so the evolution before the TAMS is mostly influenced by cluster dynamics 
(apart from the ProCeps which are secondaries), while the evolution after TAMS is mostly influenced by stellar evolution and possible mass transfer. 
The time interval from $t_{\rm 0}$ to $t_{\rm TAMS}$ is also substantially longer than 
the interval from $t_{\rm TAMS}$ to the end of core helium burning $t_{\rm He,end}$, 
so any change of orbital parameters due to 
cluster dynamics is more likely to occur between time $t_{\rm 0}$ and $t_{\rm TAMS}$ than between $t_{\rm TAMS}$ and $t_{\rm He,end}$.

For softening, we require that the orbital period of the binary either increases by at least a factor of $2$ between time $t_{\rm 0}$ and $t_{\rm TAMS}$, 
or the binary gets dissociated. 
For hardening, we require that the orbital period of the binary either decreases by at least a factor of $2$ between $t_{\rm 0}$ and $t_{\rm TAMS}$, 
or the two components have merged. 
To separate the hardening events caused by cluster dynamics (i.e. by the assistance of other cluster member(s)) from those due to 
stellar evolution, we take into account only the wider binaries (with $\log_{\rm 10} (P$[days]$) \gtrsim 3$ at $t_{\rm 0}$), 
which provides a lower estimate on the total number of hardening events.
We choose the minimum orbital period for hard binaries ($\log_{\rm 10} (P$[days]$) \gtrsim 3$) to be longer than the minimum period for a Cepheid 
with a companion on a circular orbit ($\log_{\rm 10} (P$[days]$) \approx 2.6$) to exclude the majority of the cases where the orbital period shrunk because 
of tidal circularisation during stellar binary evolution; 
only $\approx 15$ \% of the BiCeps with $\log_{\rm 10} (P$[days]$) = 3$ have eccentricity high enough ($e \gtrsim 0.67$; \citealt{Kobulnicky2014}) to 
circularise to orbital period of $\log_{\rm 10} (P$[days]$) \lesssim 2.6$.

We note that the adopted definition of soft and hard binaries differs from the classical definition of soft and hard binaries \citep{Heggie1975,Hills1975}, 
where the change of the orbital period is purely due to dynamical encounters, while the present simulations include also 
the orbital period change due to stellar evolution (which is caused by mass loss, mass transfer or the supernova kick).
We adopt this definition because the standard models do not allow us to directly distinguish  the influence of dynamical encounters and stellar evolution, 
but a comparison with control models provides an estimate of the influence of stellar evolution only.

\reff{fOrbPerChange} contrasts the outcome of binary evolution of the given initial orbital period distribution 
in star clusters (upper row) with  the control models (lower row) 
for the clusters at the galactocentric radius $R_{\rm g} = 8 \Kpc$ and metallicity $Z = 0.014$.
More than $97$\% of ProCeps are identified as being a binary member at $t_{\rm 0}$ (the remaining $3 \%$ are immediately disrupted  
by the presence of other stars in the cluster), and all the binaries are shown in the histograms. 
The outcomes in dependence of the orbital period distribution are separated by \logP{\approx 3}, where practically all 
binaries with \logP{ < 3} interact with their companion (see \refs{ssMergers} below), 
while cluster dynamics plays a more significant role for the BiCeps with \logP{ > 3}, which interact with their companions infrequently.

As expected, both hardening (magenta regions) and softening events (cyan regions) are more pronounced in models with cluster dynamics 
($p_{\rm hard} = 0.05$ and $p_{\rm soft} = 0.18$; \reft{ttestRun}) with respect to the control 
models ($p_{\rm hard} = 0.02$ and $p_{\rm soft} = 0.12$). 
The orbital period $P_{\rm sep}$ for the watershed semi-major axis $a_{\rm hard}$ \eqp{eHardThr} which separates soft binaries from hard binaries, 
ranges from $\log_{10}(P_{\rm sep} [\rm{days}]) \approx 3.2$ 
to $\log_{10}(P_{\rm sep} [\rm{days}]) \approx 6.2$ as $M_{\rm ecl}$ decreases from $6400 \Msun$ to $50 \Msun$ (grey bands in \reff{fOrbPerChange}). 
The clusters behave in a more complex way than expected from this simple picture because there are softening events for  $P < P_{\rm sep}$, 
and hardening events for $P > P_{\rm sep}$, but on average  
softening events increase with $P$, hardening events decrease with 
$P$, and hardenings are more common in less massive clusters, which have a larger value of $P_{\rm sep}$.
We note that some of the softenings and hardenings are due to stellar evolution only (lower row). 
In this case, the most of the softenings is produced by stellar mass loss of the primary, and the hardenings are produced when the former primary receives a 
supernova kick in a suitable direction. 

\reff{fperDist_v4} compares the distribution of orbital periods of BiCeps in more massive clusters (cyan line), low mass clusters (magenta), and 
in the field (black dashed line).
BiCeps in more massive clusters ($M_{\rm ecl} \gtrsim 1000 \Msun$) have shorter orbital periods (50\% of BiCeps have $\log_{10} (P$[days]$) < 4.1$) 
than BiCeps in less massive clusters (50\% of BiCeps have $\log_{10} (P$[days]$) < 5.9$ in clusters with $M_{\rm ecl} \lesssim 500 \Msun$), 
which indicates that the dynamical environment of more massive clusters preferentially destroys Cepheids with longer orbital periods. 
The orbital period distribution of BiCeps in the field lies between these two distributions, but closer to the distribution 
of lower mass clusters for longer orbital periods because lower mass clusters contribute more Cepheids to the field 
because of their higher evaporation rate and earlier disintegration.
The median orbital period of BiCeps, $Q_{\rm 50} = \rm{med}(\log_{\rm 10}(P[\rm{days}]))$, 
and its lower ($Q_{\rm 25}$) and upper ($Q_{\rm 75}$) quartile in different environments are listed in \reft{tOrbPer}. 
According to the Table, BiCeps in clusters tend to have longer orbital periods ($\rm{med}(\log_{\rm 10}(P[\rm{days}])) = 5.3$) 
than those in the field ($\rm{med}(\log_{\rm 10}(P[\rm{days}])) = 4.5$).

Companions to Cepheids with orbital periods shorter than $\approx 10$ years are of particular interest because they can be 
inferred from long-term variations of radial velocity of Cepheids. 
\citet{Shetye2024} find that $15 \pm 2$ \% of Cepheids have a companion with $\log_{10} (P$[days]$) < 3.6$, 
which is higher than in our simulations ($7$ \% for the set of models with $Z = 0.014$ and $R_{\rm g} = 8 \Kpc$)
\footnote{The present simulations do not enable us to study BiCeps with the shortest orbital periods because of a bug, which is probably at the interface 
between \nbdvi and the binary star evolution recipes of \citet{Hurley2002} because the bug is absent in studies utilising only the stellar evolutionary recipes 
(e.g. \citealt{Neilson2015a,Karczmarek2022}).
In some cases, the bug causes leaving out the common envelope phase and allows for the existence of BiCeps so close each other that their stellar radii overlap. 
For example, the shortest orbital period BiCep produced in the models with $Z = 0.014$ has $P = 53$ days. 
Accordingly, the fraction of BiCeps with $\log_{10} (P$[days]$) < 3.6$ listed above excludes companions with $\log_{10} (P$[days]$) < 2.6$.
}
.

The orbital period distribution for Cepheids in clusters and the field combined (yellow line in \reff{fperDist_v4}) has $\approx 25$\% of binaries 
with $P$ larger than the initial upper limit of the period distribution, which is $\log_{10} (P$[days]$) = 6.7$. 
These BiCeps are not primordial, but they form dynamically during the cluster evolution; they are absent in the control models (brown line). 
A closer inspection shows that the BiCeps with $\log_{10} (P$[days]$) \gtrsim 6.7$ have different origins in lower mass and more massive clusters: 
lower mass clusters form the wide BiCeps mainly during complicated interactions between the cluster stars 
and binaries throughout the cluster life-time, often 
including several consecutive exchanges; 
while more massive clusters form wide BiCeps mainly during gas expulsion at the outskirts of the clusters, with many of the 
BiCeps escaping the cluster in the process. 
The latter mechanism was studied in a general setting by \citet{Kouwenhoven2007,Kouwenhoven2010}. 
The dichotomy in the formation mechanism probably accounts for the higher ratio of $p_{\rm BC, out}/p_{\rm BC, in}$ in more massive 
clusters (left panel of \reff{fBinHiaCep}). 
Most of the long period binaries are disrupted when the Cepheid becomes a white dwarf or neutron star, with the compact object receiving a supernova kick.

\subsubsection{Dynamical exchanges}

\label{sssDynExch}

\iffigscl
\begin{figure*}
\includegraphics[width=\textwidth]{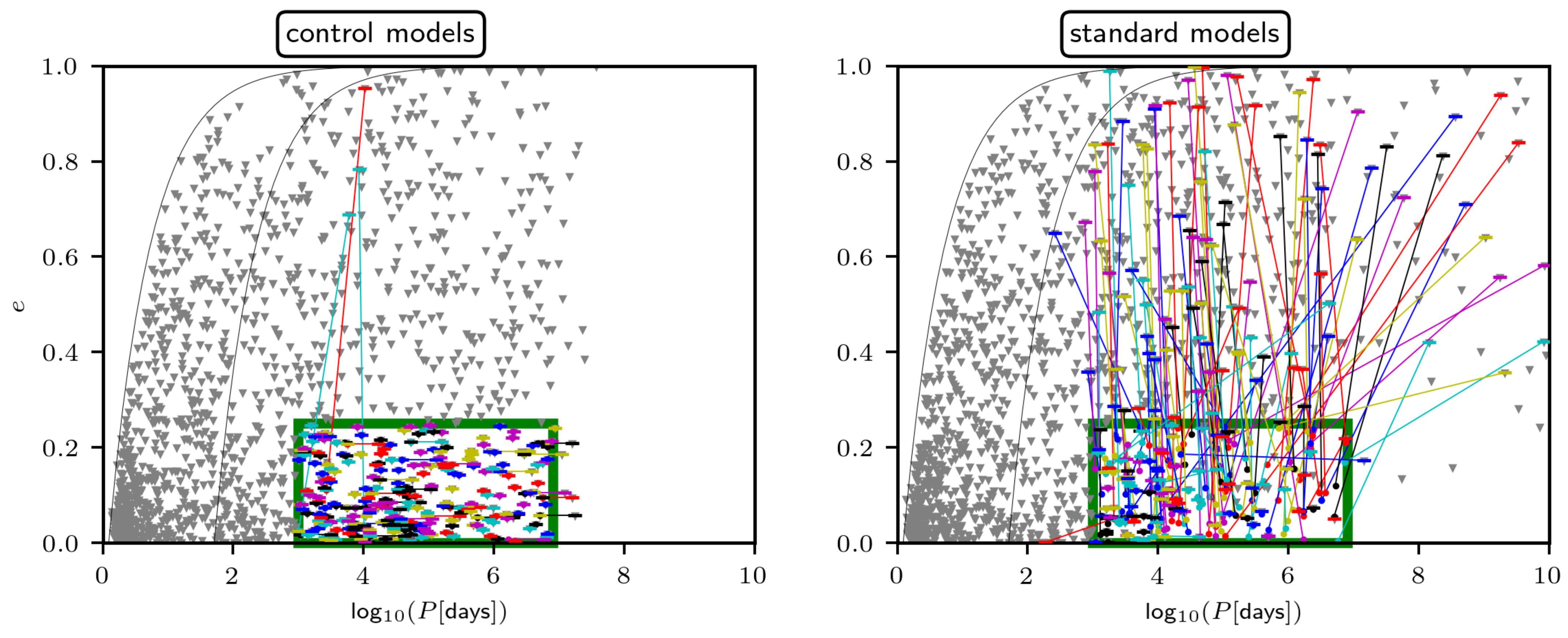}
\caption{
Evolution of the orbital period and eccentricity between the ZAMS (grey triangles) and TAMS (coloured bars) 
for the binaries which contain at least one ProCep. 
Although the plot shows all the binaries at the ZAMS, it shows the state at the TAMS only of the binaries which originate 
with $(\log_{\rm 10}(P), e)$ from the interval $(3, 7) \times (0, 0.25)$ (the green rectangle).
The lines connect the states at the ZAMS with those at the TAMS. 
The purpose of the dot and line colour is only to aid clarity. 
The black lines indicate the maximum allowed eccentricity of a binary of given orbital period to avoid a physical collision with its companion 
for $m_1 + m_2 = 9 \Msun$ and stellar radii of $R_1 = 5 \Rsun$ (on the MS) and $R_1 = 60 \Rsun$ (before the second passage through the instability strip).
\figpan{Left panel:} Control models. Only the minority of binaries evolve; 
these are the secondaries where the primary mass loss results in an increase of the orbital period. 
\figpan{Right panel:} Models with cluster environment, which result in a pronounced increase of eccentricity.
Both cases are calculated for $Z = 0.014$, $R_{\rm g} = 8 \Kpc$ and for the cluster 
mass range from $M_{\rm ecl} = 50$ to $M_{\rm ecl} = 6400 \Msun$. 
Note: the binaries which either form or are destroyed between the ZAMS and TAMS are not shown.
}
\label{fEccPlaneEvolvArr}
\end{figure*} \else \fi

\iffigscl
\begin{figure*}
\includegraphics[width=\textwidth]{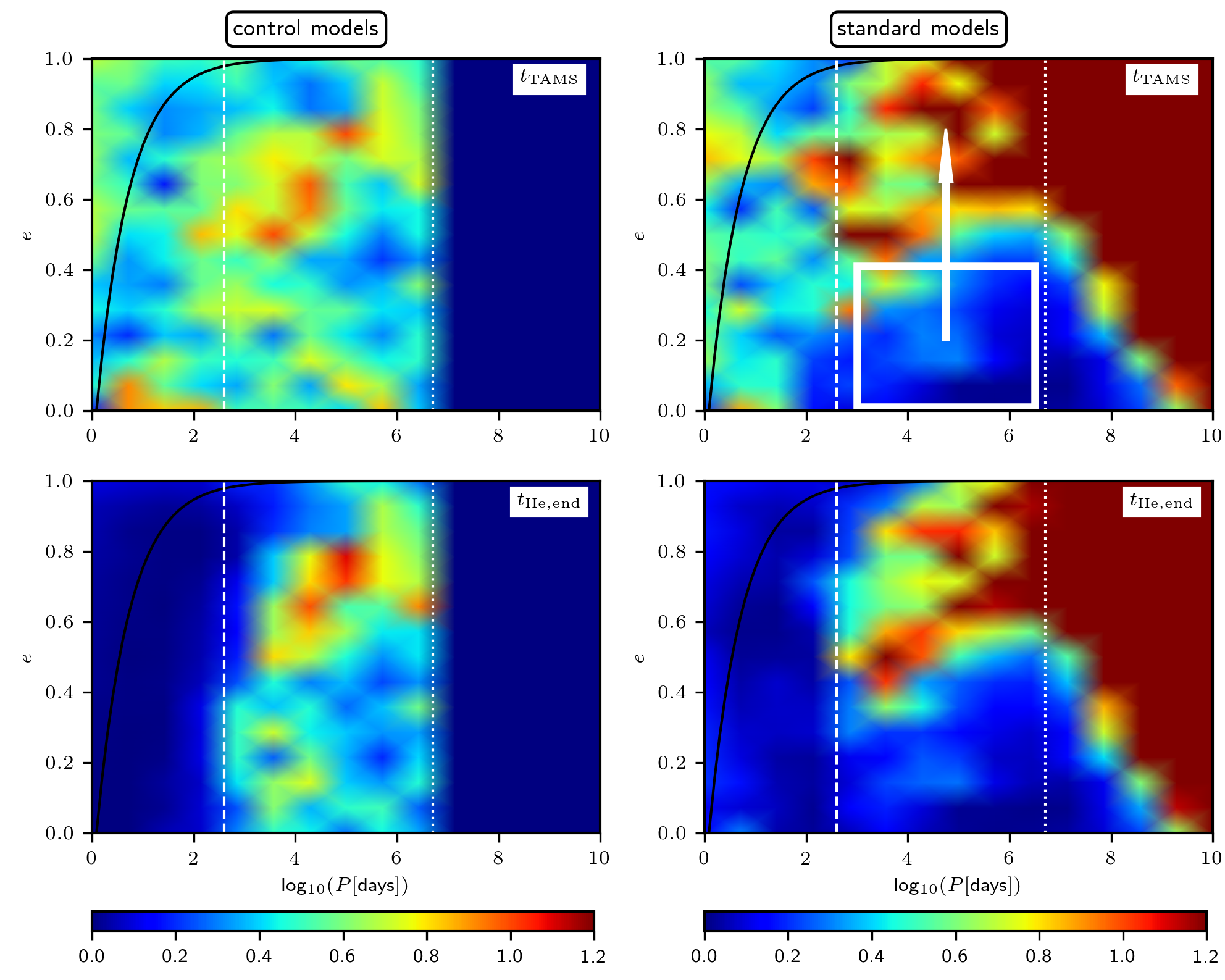}
\caption{
Change in orbital period and eccentricity between $t_{\rm 0}$ and $t_{\rm TAMS}$ (upper row), and between $t_{\rm 0}$ and the end of the core 
helium burning $t_{\rm He,end}$ (lower row). 
The colour-scale shows the ratio between the number of ProCeps (at $t_{\rm TAMS}$) or Cepheids (at $t_{\rm He,end}$) and the number of 
ProCeps at $t_{\rm 0}$, which are located at the nearby position in the $\log_{\rm 10}(P)$ - $e$ plane. 
Values larger (lower) than $1$ indicate that the number of ProCeps or Cepheids increased (decreased) since $t_{\rm 0}$. 
The dashed and dotted white lines show the period threshold for coalescence (at $\log_{\rm 10}(P[$days$]) = 2.6$) and the upper limit 
on the birth period distribution ($\log_{\rm 10}(P[$days$]) = 6.7$), respectively. 
The black line shows the maximum allowed eccentricity for given orbital period at $t_{\rm 0}$ (assuming $m_1 + m_2 = 9 \Msun$ and stellar radius of $R_1 = 5 \Rsun$).
The absence of binaries with $\log_{\rm 10}(P[$days$]) \lesssim 2.6$ at $t_{\rm He,end}$ is due to stellar collisions.
The depletion of binaries with $\log_{\rm 10}(P[$days$]) \gtrsim 3$ 
and $e \lesssim 0.5$ (the white rectangle with an indication of eccentricity evolution) 
seen already at the $t_{\rm TAMS}$ in the standard model 
is due to dynamical encounters with other cluster members (the depletion is absent in the control model). 
The binaries with $\log_{\rm 10}(P[$days$]) \gtrsim 6.7$ in the standard model (and absent in the control model) are mostly outer orbits in triples.
}
\label{fEccPlane}
\end{figure*} \else \fi

\iffigscl
\begin{figure}
\includegraphics[width=\columnwidth]{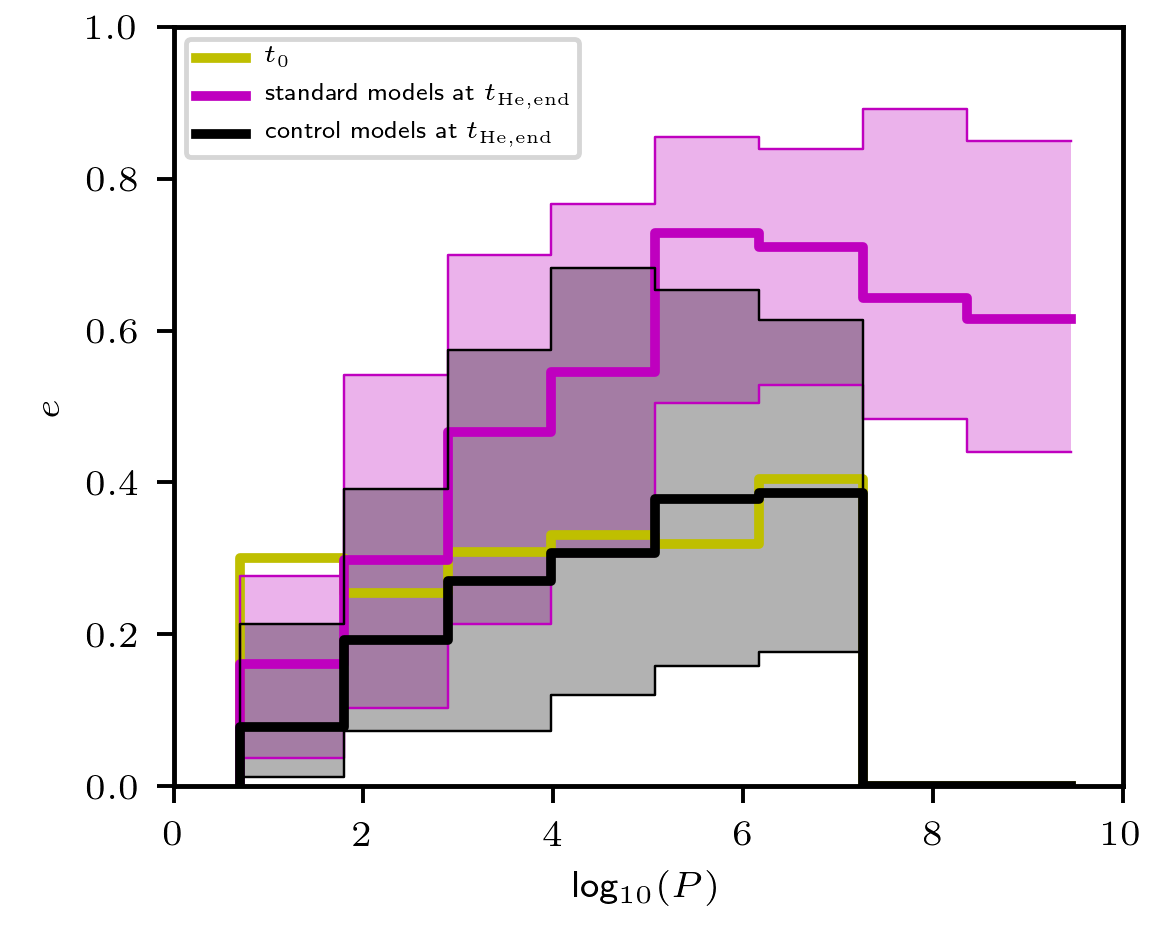}
\caption{
Eccentricity as a function of period for BiCeps in standard models (magenta) and control models (black). 
The thick lines show the mean eccentricity for given period, and the coloured areas represent the interquartile range. 
The figure quantitatively shows the boost of eccentricity for Cepheids which are subjected to cluster dynamics. 
The mean eccentricity at time $t_{\rm 0}$ (identical for both models) is shown by the yellow line.
}
\label{fmean_eccAsPer}
\end{figure} \else \fi

\iffigscl
\begin{figure*}
\includegraphics[width=\textwidth]{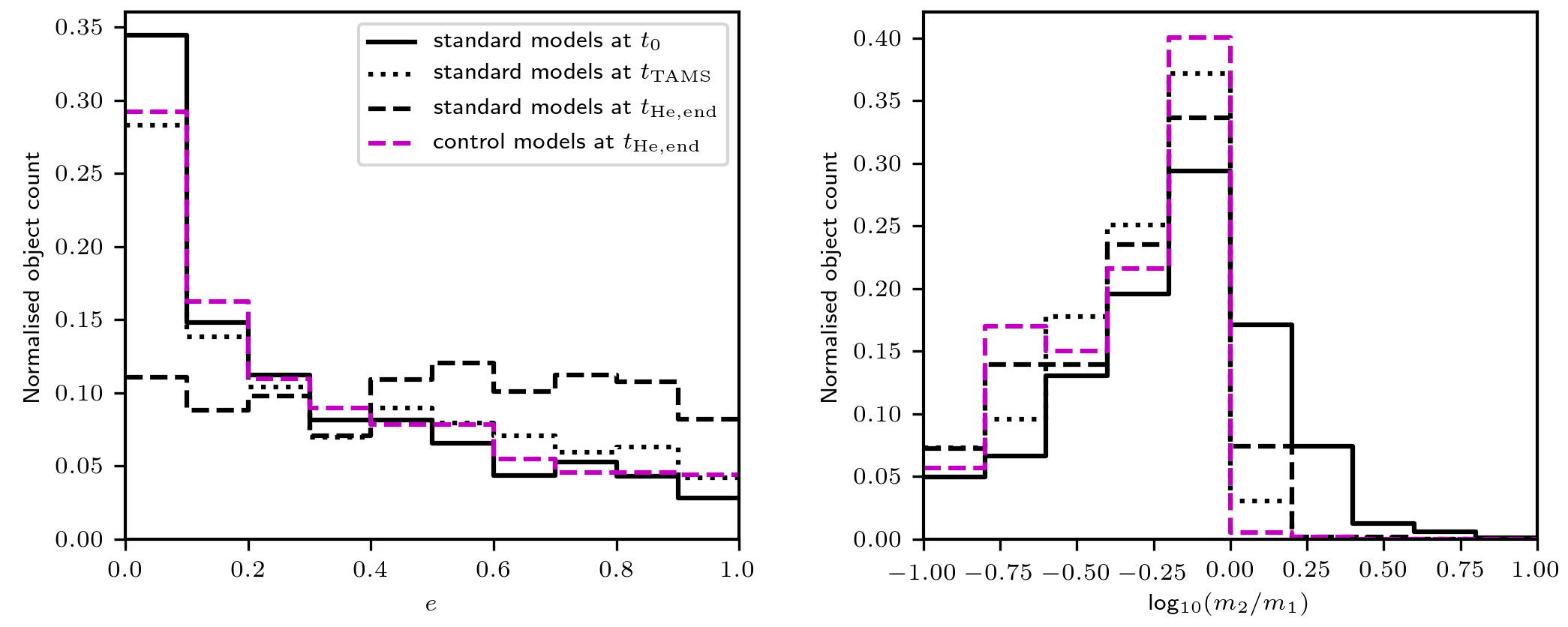}
\caption{
The eccentricity distribution (left panel) and mass ratio distribution (right panel) of ProCeps and Cepheids in binaries.
The solid, dotted and dashed line shows the distribution at $t_{\rm 0}$, $t_{\rm TAMS}$, and $t_{\rm He, end}$, respectively. 
The plot is for Cepheids formed in a population of star clusters with $R_{\rm g} = 8 \Kpc$, $Z = 0.014$, and of the adopted ECMF of slope $\beta = 2$ \eqp{eCSP}.
}
\label{fperDist_v3}
\end{figure*} \else \fi

Although some Cepheids are in binaries, their companions are not always the same stars with which the Cepheids formed. 
This happened due to a dynamical exchange involving at least three stars. 
Alternatively, some of these BiCeps originate from the primordial binaries which capture an outer body forming a temporal triple, 
and where the primordial (inner) binary merges later. 
The outer body becomes a new companion to the merged object, with the possibility of
the Kozai-Lidov mechanism triggering some of the coalescences \citep{Perets2009,Naoz2014}; in both cases, the identity of the companion has changed. 
The fraction $p_{\rm chg}$ of BiCeps which have exchanged their companion between time $t_{\rm 0}$ and TAMS is $\approx 0.10$ (\reft{ttestRun}). 
This quantity increases with decreasing cluster mass from $\approx 0.05$ for the $6400 \Msun$ clusters 
to $\approx 0.20$ for the $50 \Msun$ clusters. 
Exchanges occur in BiCeps over the whole orbital period range with a mild preference for longer orbital periods 
(the yellow area in the upper row of panels in \reff{fOrbPerChange}).

\subsubsection{Cepheids with compact companions}

\label{sssCompactComp}

Next, we study the fraction of Cepheids $p_{\rm cmp}$ which have a compact companion. 
This estimate is subject to large uncertainties because it depends not only on the details of orbital period changes during possible Roche-lobe  
mass transfer, but also on the supernova kick velocity distribution, which is not well-known \citep[e.g.][]{Hansen1997,Hobbs2005,Fragos2009,Beniamini2016,Giacobbo2020}. 
Assuming that the supernova kicks follow a Maxwell distribution with dispersion $\sigma = 40 \Kms$, 
$\approx 3$ to $5$\% of all Cepheids have a compact companion (\reft{ttestRun}). 
The compact companion is a white dwarf in $\approx 80$\% of the cases, a neutron star in $\approx 16$\% of the cases and a black hole in 
$\approx 4$\% of the cases.
The orbital period distribution of Cepheids with compact companions is more narrow (interquartile range of $Q_{\rm 75} - Q_{\rm 25} = 1.4$; \reft{tOrbPer}; 
the dotted line in \reff{fperDist_v4}) 
than for Cepheids with non-degenerate companions with practically no binaries of \logP{\gtrsim 7}.

As expected, BiCeps with compact companions have a smaller mass ratio than the BiCeps with gaseous companions; 
the median and interquartile range is $\log_{\rm 10} (m_2/m_1)$ of $-0.67$ and $0.07$ for the former, and 
$-0.32$ and $0.54$ for the latter. 
The mass ratio is significantly more narrow for compact companions because of the relatively small mass 
range allowed for neutron stars and white dwarfs originating from stars of mass $\gtrsim 4 \Msun$.

\subsubsection{Cepheids originating from secondaries}

In the present model of binary stars at metallicity $Z = 0.014$, approximately 26 \% of ProCeps formed as secondaries (the remaining 74 \% of ProCeps formed as primaries).
The ProCeps which formed as secondaries are more impacted (by collisions or mass transfer) with their more massive and faster evolving companions than the 
ProCeps which formed as primaries: 
Only $38$ \% of the ProCeps which formed as secondaries become Cepheids, while $80$ \% of the ProCeps which formed as primaries become Cepheids. 
Among all RanCeps, only 14 \% were initially secondary ProCeps (the rest were primaries).
The RanCeps which originated from secondaries have a substantially smaller binary fraction ($p_{\rm BC} \approx 0.15$) 
than the RanCeps which originated from primaries ($p_{\rm BC} \approx 0.5$). 
The lower binary fraction for the secondary Cepheids is likely caused by the supernova kick to the primaries 
and the dynamical disruption of the binary as described in the paragraph below.
The majority of the binary RanCeps which were initially secondaries have 
a compact companion: $\approx 30$ \% of their companions are neutron stars,  
$\approx 50$ \% white dwarfs, and only 
$\approx 20$ \% gaseous stars, which are typically also Cepheids or red giants with a slightly larger mass than the secondary. 

Apart from stellar evolution in binaries, the Cepheids which formed as secondaries are more influenced by stellar dynamics, 
usually in the form of the binary-binary or similar interaction between stellar subsystems, which gives the former secondaries 
enough speed to escape their birth clusters and possibly releases them from their companions. 
The ejected stars travel at larger speeds and thus they can be found at larger distances from the clusters. 
We take the cut-off distance to be $300 \Pc$ because it is the distance travelled by a star at a speed corresponding 
to the velocity dispersion of a $500 \Msun$ cluster ($2 \Kms$) for the $150 \Myr$ life-span of Cepheids at the Solar metallicity. 
While $34$ \% of the Cepheids which were initially secondaries are located farther away than $300 \Pc$ from their birth clusters, 
this fraction is only $11$ \% for the Cepheids which were initially primaries.
This behaviour is expected from N-body dynamics where binary-binary or binary-single star interactions preferentially release the least massive 
stars from the subsystem, i.e. the secondaries \citep{Valtonen1974,Heggie2003,Tanikawa2012}.


\subsubsection{Binaries where both the primary and secondary are Cepheids}

\label{sssBothCeps}

Some Cepheids have as a companion another Cepheid. 
To calculate their fraction $p_{\rm C-C}$ among all Cepheids, we divide the time interval of $(0, 300 \Myr)$ to bins of $30 \Kyr$, 
which are sufficiently short to resolve the movement of the Cepheid in the instability strip 
for the most massive Cepheids (they have the fastest evolution), and we count the  
Cepheids, $N_{\rm C,i}$, in each bin $i$ as well as the Cepheids, $N_{\rm C-C,i}$, in a binary with another Cepheid. 
The definition $p_{\rm C-C} = \sum_{\rm i} N_{\rm C-C,i}/\sum_{\rm i} N_{\rm C,i}$, where $i$ runs over all time bins, 
provides the expected observed fraction of Cepheids residing in a binary of two Cepheids when averaged over Cepheids of all possible ages 
forming within a galaxy of a time-independent star formation rate. 
Both standard and control models provide $p_{\rm C-C} = 0.013$ with median orbital periods \logP{\approx 4.6}, 
which is slightly shorter than the orbital period distribution of all Cepheids (\reft{tOrbPer}).
The small stock of statistics available does not allow for the metallicity dependence of this quantity to be properly studied.
We note that \citet{Karczmarek2022} found lower values of $p_{\rm C-C} \approx 0.001$ to $0.01$ for most of their binary models.

\subsubsection{Evolution of eccentricity}

Star cluster dynamics is responsible for substantial changes in eccentricity. 
The lines in \reff{fEccPlaneEvolvArr} connect the states in the orbital period-eccentricity plane ($\log_{\rm 10}(P)$ - $e$) 
of the binaries containing at least one ProCep at the ZAMS with their states at the TAMS. 
The comparison is done at the TAMS because the time-span covers the majority of the stellar life-time ($\approx 90$\%) so that 
cluster dynamics has caused most of its evolution of orbital parameters while stellar evolution has still relatively minor influence.
For the sake of clarity, the Figure shows 
the evolution only of the binaries having $3 < \log_{\rm 10}(P\rm{[days]}) < 7$ and $e < 0.25$ at the ZAMS. 

In the control models (left panel), only a small fraction of binaries undergo evolution of their orbital 
parameters between the ZAMS and TAMS. 
Most of the binaries that evolve their orbital parameters increase their orbital periods at constant eccentricity, 
which is due to the stellar mass loss from the primary 
in the binaries where the Cepheid is the secondary. 
In rare cases, some binaries increase their eccentricity, which is due to the SN explosion and kick in the primaries where the Cepheid is the secondary.

In contrast, cluster dynamics causes more significant changes of orbital parameters (right panel of Figure \ref{fEccPlaneEvolvArr}), 
particularly an increase of eccentricity. 
For example, in the control models practically all BiCeps with $e > 0.75$ had already $e > 0.75$ at the ZAMS (this is $11$\% of all BiCeps), 
while in the standard models cluster dynamics increases the fraction of BiCeps with $e > 0.75$ to $23$\%. 
Moreover, cluster dynamics also dissolves about half of the initially highly eccentric binaries so that 
only approximately one in four of the highly eccentric BiCeps had $e > 0.75$ 
already at the ZAMS ($6$\% of all future BiCeps have $e > 0.75$ at the ZAMS), 
the rest being produced by encounters in the cluster.

The ratio between the number of ProCeps in binaries at $t_{\rm TAMS}$ and $t_{\rm 0}$ at a given area in the $\log_{\rm 10}(P)$ - $e$ plane 
is shown in \reff{fEccPlane}
\footnote{The density of binaries is calculated on the $\log_{\rm 10}(P)$ - $e$ grid by the method analogous to that of \citet{Casertano1985}. 
The only difference is that we take into account $30$ binaries to reduce noise and that we work in a two dimensional plane instead of 
a three dimensional space. 
The local density of binaries at $t_{\rm TAMS}$ is then divided by the local density at $t_{\rm 0}$ to obtain \reff{fEccPlane}.}
.
Between these two time events, the interaction of the binaries with other cluster members reduces the population of low eccentric 
binaries ($e < 0.25$) with $3 < \log_{\rm 10}(P\rm{[days]}) < 7$ to $27$\% (upper right panel), 
which is a substantially larger depletion than due to stellar evolution only (population reduced to $70$\%; upper left panel). 
Cluster environment also forms binaries of longer orbital periods ($\log_{\rm 10}(P\rm{[days]}) > 7$) than present at $t_{\rm 0}$.

When the ProCeps become Cepheids, cluster environment leaves its imprints in the 
eccentricity of Cepheids with $\log_{\rm 10}(P\rm{[days]}) \gtrsim 3.5$ (\reff{fmean_eccAsPer}). 
For example, Cepheids with $\log_{\rm 10}(P\rm{[days]}) \approx 6$ have  
median eccentricity of $0.38 \pm 0.25$ (the error represents the interquartile range) in the control models, 
while they have median eccentricity of $0.73 \pm 0.16$ in the standard models.
Future observations of eccentricity distribution among Cepheids with wide companions might 
help constrain the environment in which they were born.

Interestingly, albeit the depopulation of the low eccentricity binaries with $3 < \log_{\rm 10}(P\rm{[days]}) < 7$ is striking 
and it practically leads to a desert (upper right panel of \reff{fEccPlane}), this fact is not so apparent from the eccentricity distribution 
of all ProCeps as indicated by the dotted histogram in the left panel of \reff{fperDist_v3}. 
This is because the population of low eccentric binaries is dominated by the binaries 
of short orbital periods (\logP{\lesssim 2.5}), whose dynamical 
evolution is less significant because of their compactness and thus smaller cross sections for interaction \citep{Hut1983}. 
As the ProCeps evolve beyond the TAMS, their radii increase, and they interact with their companions. 
In this process, practically all ProCeps with \logP{\lesssim 2.5} interact with their companion so that they are either prevented from becoming Cepheids, 
or they are single stars when becoming Cepheids; in any case the states with \logP{\lesssim 2.5} are erased as the stars evolve from 
the TAMS to $t_{\rm He, end}$ (lower row of \reff{fEccPlane}). 

Since the close binaries are predominantly the ones which have low eccentricity at the TAMS, 
the eccentricity distribution changes between the $t_{\rm TAMS}$ and $t_{\rm He, end}$ for the standard models substantially. 
While the eccentricity distribution becomes almost flat in the standard models, it is only marginally modified in the control models between 
$t_{\rm 0}$ and the Cepheid stage (cf. the dashed lines in the left panel of \reff{fperDist_v3}), 
implying that binary evolution only is not able to change the eccentricity distribution to a larger extent. 
Thus, the desert for \logP{\gtrsim 3} and $e \lesssim 0.5$ and the flattening of eccentricity distribution is the result of cluster dynamics.

\subsubsection{Evolution of the mass ratio}

Cluster evolution does not noticeably influence the mass ratio distribution between the BiCep and its companion 
(right panel of \reff{fperDist_v3}). 
The median mass ratio $\log_{\rm 10} (m_2/m_1)$ decreases from $-0.16$ for ProCeps at $t_{\rm 0}$ to $-0.32$ for Cepheids in the models including cluster 
dynamics (for models with $R_{\rm g} = 8 \Kpc$, $Z = 0.014$) and to $-0.29$ in control models (for models with $Z = 0.014$).

\subsection{Cepheids in triples and quadruples}

\label{ssTriQ}

\iffigscl
\begin{figure*}
\includegraphics[width=\textwidth]{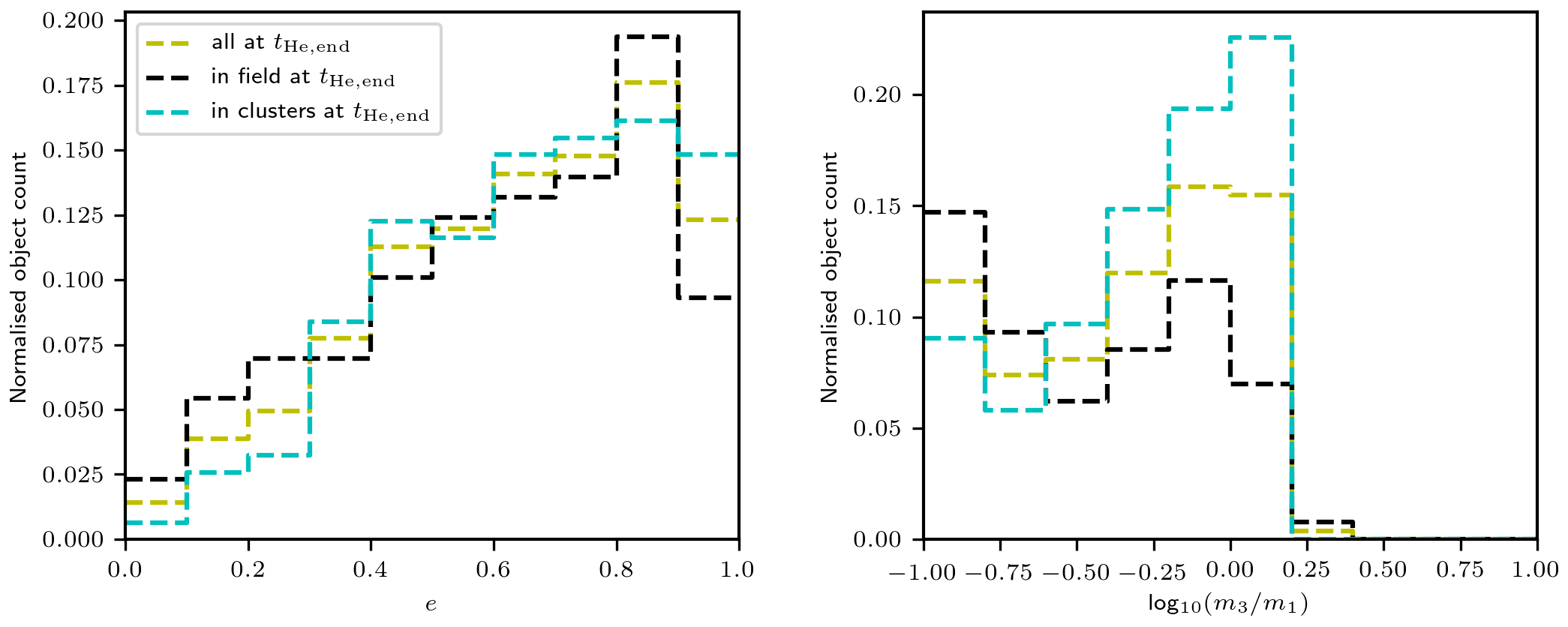}
\caption{
Eccentricity distribution (left panel) and mass ratio distribution (right panel) of the outer bodies in triples including Cepheids. 
The distribution for all triples (i.e. in star clusters and in the field combined) 
and the triples in the field and clusters only is shown by the yellow, black and cyan lines, respectively.
We note that both distributions are remarkably different from the distributions of Cepheids in binaries, which are shown in \reff{fperDist_v3}.
}
\label{fperDistTriples}
\end{figure*} \else \fi

\iffigscl
\begin{figure}
\includegraphics[width=\columnwidth]{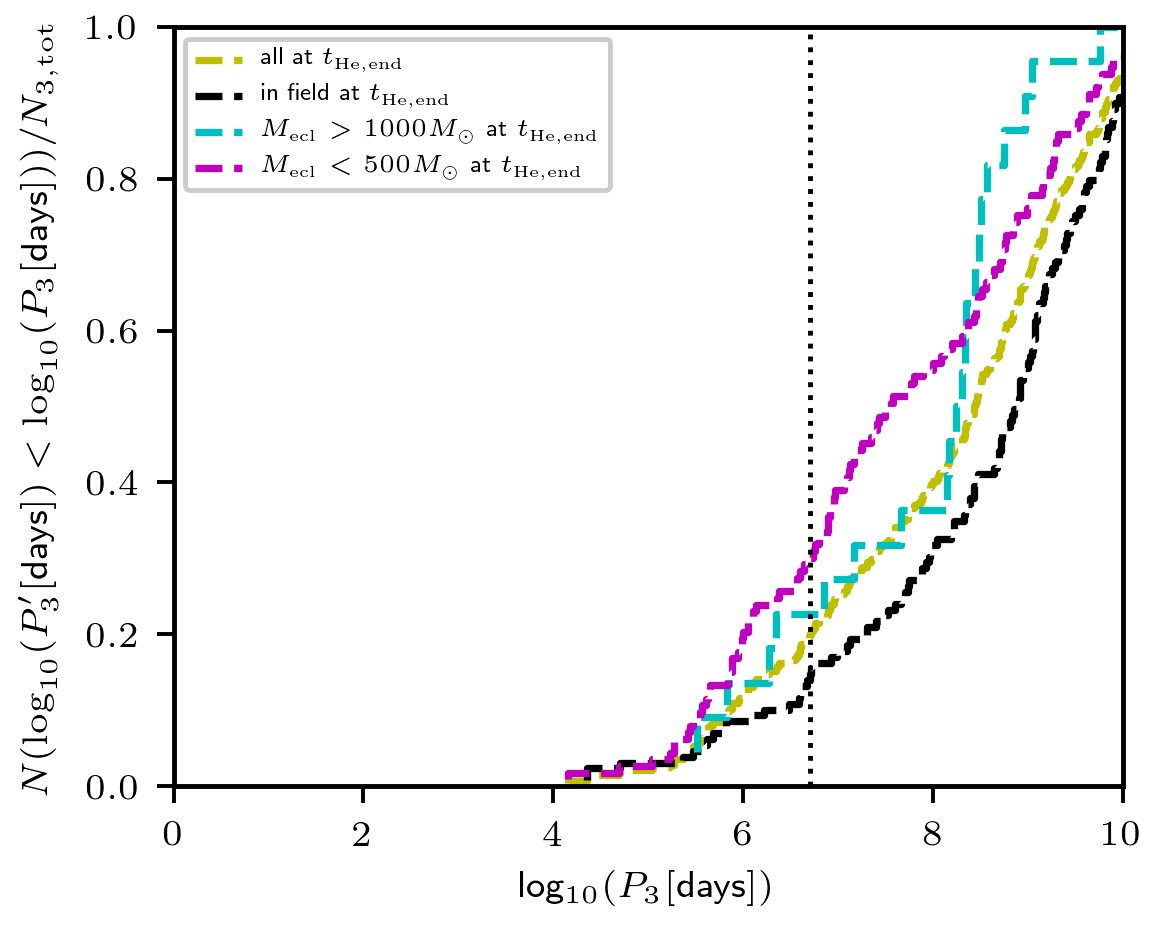}
\caption{
Cumulative distribution of orbital periods of the outer orbits in triples including Cepheids. 
We show separately the distribution for all Cepheids in triples in clusters and field combined (yellow lines), 
only in field (black lines), the Cepheids originating from more massive clusters (cyan), and the Cepheids originating from lower mass clusters (magenta).
The distribution is normalised to the total number of triples $N_{\rm 3,tot}$ in each group.
}
\label{fperDist_v5}
\end{figure} \else \fi

\iffigscl
\begin{figure}
\includegraphics[width=\columnwidth]{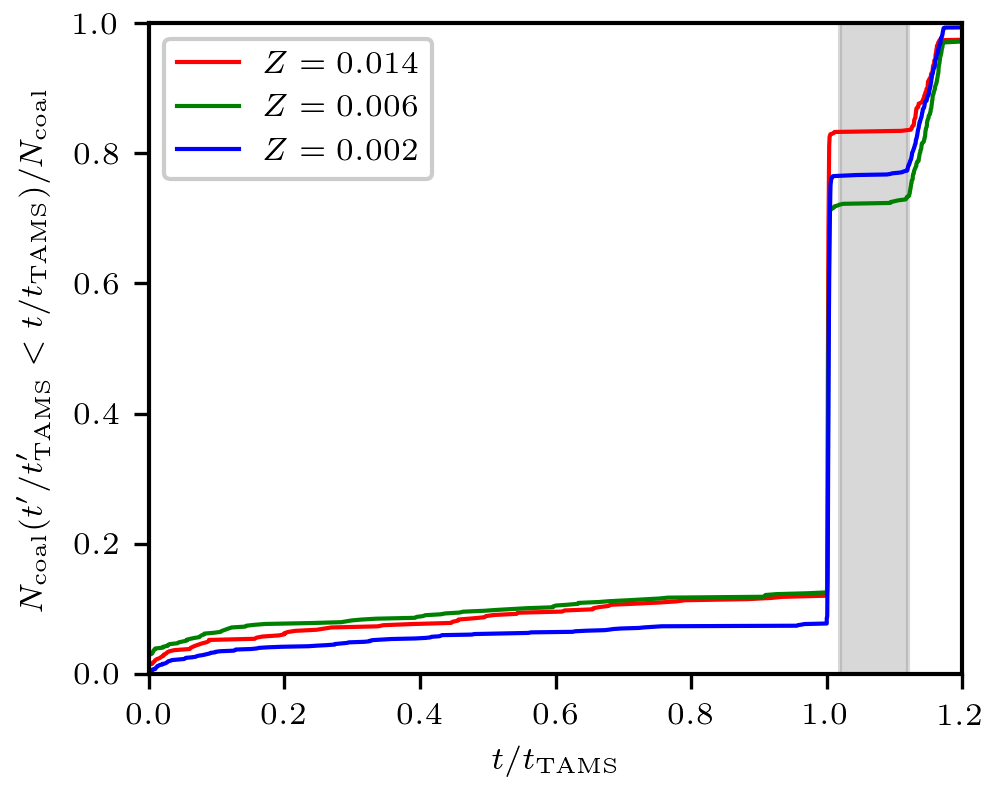}
\caption{
Cumulative distribution of the time of coalescence for the RanCeps that experienced a coalescence normalised to the $t_{\rm TAMS}$ of each individual Cepheid. 
Results for clusters with $R_{\rm g} = 8 \Kpc$ are plotted separately for the metallicity $Z = 0.014$ (red line), $Z = 0.006$ (green) and $Z = 0.002$ (blue). 
The approximate time when the stars are located within the Cepheid instability strip is indicated by the grey area.
}
\label{ftcoal}
\end{figure} \else \fi

We adopt the definition of the fraction of Cepheids 
in triples $p_{\rm tr}$ as the ratio of all Cepheids with at least two companions to the total number of Cepheids.
As with binaries, we include only the systems for which the outer body has the orbital period around the inner binary shorter than $10^{10}$ days.
The fraction of Cepheids in triples as a function of the initial 
mass of their birth cluster $M_{\rm ecl}$ is shown in the right panel of \reff{fBinHiaCep} 
separately for Cepheids in clusters ($p_{\rm tr, in}$; solid lines) and in the field ($p_{\rm tr, out}$; dashed lines).
The plots provide a lower limit of Cepheids in triples because present models do not include primordial triples; all 
triples form dynamically. 
The triple fraction is larger in clusters ($p_{\rm tr, in} = 0.31$) than in the field ($p_{\rm tr, out} = 0.11$), 
and $p_{\rm tr,\{in,out\}}$ decreases with increasing $M_{\rm ecl}$ 
with only a slight dependence on $R_{\rm g}$ or $Z$.
This is similar to the trend seen in binaries (the binary fraction is $1.7 \times$ higher in cluster than in the field), 
albeit it is more pronounced in triples (the triple fraction is $3 \times$ higher in cluster than in the field).
The triple fraction of Cepheids in clusters and field combined is $p_{\rm tr} \approx 0.18$.

Outer orbits (i.e. the orbit of the outer star around the inner binary) have nearly thermal eccentricity distribution 
(left panel of \reff{fperDistTriples}, where $f(e) \propto e$; \citealt{Heggie1975}), 
which is consistent with their formation due to capture. 
This distribution stands in contrast to the flat eccentricity distribution of binary Cepheids (left panel of \reff{fperDist_v3}). 
Another difference between the distribution of binaries and outer orbits in triples is in the mass ratio distribution,
which is more flat and bimodal with peaks at $\log_{\rm 10} (m_3/m_1) \approx -0.9$ and $\approx -0.1$ (right panel of \reff{fperDistTriples}), 
while the distribution of binaries increases with $\log_{\rm 10} (m_2/m_1)$ (right panel of \reff{fperDist_v3}). 
Triples in clusters tend to have outer companions of more equal masses than triples in the field (c.f. the cyan and black lines 
in the right panel of \reff{fperDistTriples}).
The mass ratio distribution for triples is calculated as the ratio of the outer body of mass $m_3$ relatively to the Cepheid, 
which is of mass $m_1$.
In less frequent cases, where a single Cepheid is being orbited by a binary, $m_3$ is taken to be the mass of the binary.

The period distribution of outer orbits is shown in \reff{fperDist_v5} separately for Cepheids in different environments: for all Cepheids (yellow line), 
all Cepheids in field (black line), Cepheids in more massive clusters ($M_{\rm ecl} > 1000 \Msun$; cyan line) and Cepheids 
in lower mass clusters ($M_{\rm ecl} < 500 \Msun$; magenta line). 
In contrast to BiCeps (\reff{fperDist_v4}), the distribution does not depend sensitively on the environment. 
Outer orbits also have much longer orbital periods with median $\log_{\rm 10}(P$[days]$)$ of $8.5$ and 
interquartile range of $Q_{\rm 75} - Q_{\rm 25} = 2.2$ (\reft{tOrbPer}), 
nevertheless some outer orbits have orbital periods comparable to that of BiCeps ($\approx 20$\% of outer orbits have \logP{\lesssim 6.7}). 
Figures \ref{fperDistTriples} and \ref{fperDist_v5} are drawn for Cepheids formed by a population 
of star clusters with $R_{\rm g} = 8 \Kpc$, $Z = 0.014$, and of the adopted ECMF of slope $\beta = 2$ (eq. \ref{eCSP}).

The fractions of Cepheids in quadruples in clusters $p_{\rm quad,in}$ and in the field $p_{\rm quad,out}$ follow similar trends as binaries and triples: 
lower mass clusters have a significantly larger fraction of quadruples ($p_{\rm quad,in} \approx 0.10$ and $p_{\rm quad,out} \approx 0.05$) than more massive clusters, 
which have $p_{\rm quad,in}, p_{\rm quad,out} \lesssim 0.01$. 
The combined quadruple fraction of Cepheids in clusters and in the field is $p_{\rm quad} \approx 0.04$.
Given the low number statistics of quadruples in our simulations, we do not detect 
the dependence of $p_{\rm quad,in}$ and $p_{\rm quad,out}$ on $R_{\rm g}$ or $Z$. 
For a particular cluster mass, the absolute number of quadruples is lower by a factor of $5$ than the number of triples.

All triples in our simulations form dynamically. 
As we describe in sect. 4.2 of Paper I, in lower mass clusters ProCeps mass segregate to the cluster centre before the cluster density 
is diluted as the result of gas expulsion. 
Their elevated density and relatively low velocity near the cluster centre enables binary formation where one or 
both of the inner components of the binary is another binary, 
so a triple or quadruple forms dynamically \citep{Binney2008,Moeckel2010}. 
This is the same mechanism we discuss in the context of binary formation at the end of \refs{sssChangesPeriods}.

An example of dynamical formation of a stellar multiplet that includes several ProCeps is shown in the online material. 
The movie shows a $100 \Msun$ cluster (model M7), which mass segregates its ProCeps before gas expulsion, 
so that they are retained near the cluster centre, where they survive the potential rebound due to gas expulsion (see Sect. 4.2 of Paper I for details).
Located in a small volume at the cluster centre, all the ProCeps in the clusters rapidly (by $1.1 \Myr$) form a hierarchical sextuplet.
The sextuplet ejects a binary containing one of the Cepheids at $\approx 3.6 \Myr$. 
During the ejection, the new quadruple (consisting of two hard binaries orbiting each other) captures three low mass stars,
forming a hierarchical septuplet. 
All the captured stars have eccentric orbits. 
The three captured stars can be seen in the movie (upper right panel) after the ejection event. 

The formation of hierarchies is preferred in lower mass clusters because they contain a small number of stars of mass similar to ProCeps and these 
stars mass segregate rapidly to the cluster centre, which shortens the timescale for multiple star formation \citep[][their sect. 7.1]{Binney2008}. 
The multiple star system does not need to be inherently stable, but its decay-time (which, for unstable systems, corresponds to $\approx 10$ to $100$ 
crossing times; \citealt{Heggie2003,Wang2019}) is relatively long (even comparable to $t_{\rm He,end}$ in some cases) so that the Cepheid might be 
still in a multiple stellar system. 
This explains why the fraction of Cepheids in triples increases with decreasing cluster mass. 
Triples also need time to be formed dynamically, which explains why field Cepheids (most of which escape when the clusters are young) 
have a lower triple fraction than the ones in clusters. 
Another way how to transport triples to the field is to eject them; however, ejecting a triple out of the cluster is not easy as 
these systems can be disrupted in the encounter which would eject it. 

\subsection{Stellar mergers and mass transfer events}

\label{ssMergers}

We define the number fraction of ExCeps 
(i.e. stars which are prevented from becoming Cepheids by a merger, or Roche-lobe mass transfer, or common envelope evolution with their companion)
as $p_{\rm EC} \equiv 1 - N_{\rm RC}/N_{\rm PC}$, where $N_{\rm RC}$ is the total number of RanCeps 
and $N_{\rm PC}$ is the total number of ProCeps at time $0$.
The value of $p_{\rm EC}$ is independent of the Galactic radius $R_{\rm g}$ ($p_{\rm EC} \approx 0.30$ for $Z = 0.014$; \reft{ttestRun}) 
with a possible weak decrease with decreasing metallicity ($p_{\rm EC} \approx 0.27$ for $Z = 0.002$). 
These fractions are in good agreement with the estimate $p_{\rm EC} = 0.25 \pm 0.15$ provided by
\citet[][their sect. 6.1]{Moe2017}.
Cluster dynamics increases $p_{\rm EC}$ by approximately $20$\% (cf. with the control models in \reft{ttestRun}). 
The vast majority of ExCeps ultimately merge with their companion; only $\lesssim 4$ \% of ExCeps are caused solely by mass transfer without a merger event.

\iffigscl
\begin{figure}
\includegraphics[width=\columnwidth]{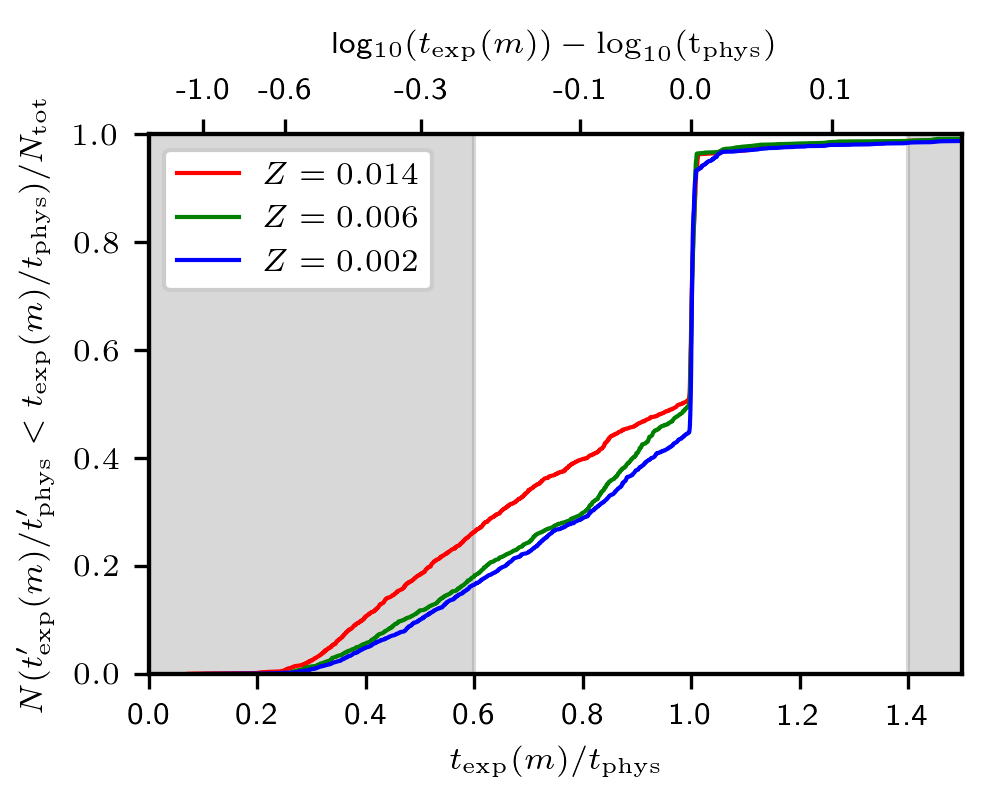}
\caption{
Cumulative distribution of the ratio of the Cepheid age $t_{\rm exp} (m)$ as expected from its present mass 
to its physical age $t_{\rm phys}$ (i.e. the time since its formation) for all Cepheids which are located in star clusters. 
Cluster metallicity is shown by colour.
The grey regions indicate the cases where $t_{\rm exp} (m)$ differs by more than $40$\% from $t_{\rm phys}$. 
These cases, whose fraction among all Cepheids is denoted $p_{\rm mma}$, are likely to be detected by observers as Cepheids of discrepant ages. 
}
\label{ftmmaRatio}
\end{figure} \else \fi

While some mergers with companions 
prevent ProCeps from becoming Cepheids, other mergers still produce Cepheids. 
The fraction of Cepheids which have experienced a merger (i.e. the total number of Cepheids which have merged divided by the total number of Cepheids) 
and which are located in clusters ($p_{\rm C,coll,in}$) and in the field ($p_{\rm C,coll,out}$) are listed separately in \reft{ttestRun}.
The Cepheids located in clusters have merged more often than the Cepheids in the field (e.g. $p_{\rm C,coll,in} = 0.45$ and 
$p_{\rm C,coll,out} = 0.36$ for $R_{\rm g} = 8 \Kpc$ and $Z=0.014$).
The difference is probably caused by the hardening events, which are possible only within star clusters. 
We do not detect a significant dependence of $p_{\rm C,coll,in}$ and $p_{\rm C,coll,out}$ on $R_{\rm g}$ or $Z$, except for a possible slight increase for $Z = 0.002$.

At which evolutionary stage is the ProCep most likely to merge with its companion? 
\reff{ftcoal} shows the cumulative histogram of all Cepheids which have merged with their companion before time $t$ normalised to $t_{\rm TAMS}$ 
for each Cepheid. 
Approximately $12$\% of mergers occur still on the MS for $Z = 0.014$; the ratio decreases in lower metallicity environments, which is probably 
because of the lower mass threshold for Cepheids (i.e. they have smaller radius) and generally smaller stellar radius at a lower metallicity 
for given stellar mass on the MS \citep{Tout1996}, both decreasing the closest distance between stars for an interaction. 
The majority of mergers ($60$\% to $70$\%) occur shortly after $t_{\rm TAMS}$ when the star substantially increases its radius. 
Approximately $20$\% of mergers occur after the Cepheid stage (i.e. after $t_{\rm He,end}$), which has no influence on the Cepheid.

Some ProCeps in clusters have merged with wide companions (\logP{> 4}; upper row of \reff{fOrbPerChange}) even though such 
cases are excluded from purely binary evolution because even moderately eccentric ($e \approx 0.5$) 
massive primaries ($m \gtrsim 15 \Msun$) can interact with their companions 
during their red supergiant phase only when \logP{\lesssim 3.8} \citep{Bertelli2009,Moe2017}. 
These cases are absent in our control models (lower row of \reff{fOrbPerChange}). 
We suspect that these events originate from the Kozai-Lidov mechanism \citep{Perets2009,Naoz2014}, where the initial relatively 
mild eccentricity of the inner binary was boosted by the outer component in a triple. 
The outer component was captured during the cluster evolution before the star became Cepheid.  
This idea is supported by the fact that $\approx 93$\% of the Cepheids which have \logP{> 4} at $t_{\rm 0}$ and 
which have merged with their companions have an outer companion at $t_{\rm He,end}$ (this companion might have caused the merger via the Kozai-Lidov mechanism), 
while the percentage of the Cepheids having \logP{< 4} at $t_{\rm 0}$ and a distant companion is only $42$\%.


The Cepheids which have experienced stellar mergers or Roche-lobe mass transfer in their history appear in their host clusters at a different time (usually later) 
than what is expected for their mass (pulsation period) and the cluster age, which makes them peculiar objects.
\reff{ftmmaRatio} shows the cumulative histogram of the ratio between the expected age $t_{\rm exp} (m)$ of the Cepheid calculated from 
its mass to its physical age $t_{\rm phys}$ (taken as the time since $t_{\rm 0}$). 
The age of a substantial number of Cepheids ($\approx 50$\%) has been influenced by their interaction with their companion, 
with some of them ($\approx 3$\%) having $t_{\rm exp} (m)$ shorter by more than a factor of $3$ than their physical age. 
This opens the interesting possibility for the star clusters which host multiple Cepheids that one or several of 
the Cepheids could appear to be of different age than the other Cepheids within the same cluster even though they actually formed in a coeval star forming event. 
These Cepheids, whose age can be estimated from period-age relations \citep{Anderson2016a,DeSomma2021}, 
would not match the age of the cluster estimated from isochrone fitting using the same set of stellar evolutionary models.
Therefore, Cepheids in clusters might be imprecise indicators of the cluster age, often underestimating it.
A merger which appears younger than the rest of the cluster and which is likely to evolve to a Cepheid in the future 
has already been observed in the Pleiades (star Alcyone of mass $\approx 6.5 \Msun$; \citealt{Brandt2015}).

A small fraction ($\approx 3$\%) of Cepheids appear to be older by more than $10$\% than expected from their mass (see \reff{ftmmaRatio}), 
i.e. these stars are of too low mass for their age. 
Most of these cases had been eccentric binaries before their evolution out of the MS, whereupon they interacted with their companion 
in such a way that resulted in some mass loss and tidal circularisation; nevertheless the secondary still provided the primary with enough room 
to become a Cepheid. 
These binaries usually coalesce immediately after the Cepheid stage. 
Apart from this evolutionary path, some of the Cepheids with $t_{\rm exp} (m)/t_{\rm phys} > 1.1$ originate from mergers of stars with very 
high angular momentum, which lose more mass in the merger than what they gain from the secondary because of angular momentum conservation. 

In total, from 15\% to 30\% of all Cepheids found in clusters have either coalesced with their secondary or gained significant amount of mass from their former primary
so that their derived age is substantially different (by more than $40$\%) from the age of the cluster. 
The fraction of these cases is labelled $p_{\rm mma}$ in \reft{ttestRun}, and they are represented by the grey regions of \reff{ftmmaRatio}. 
We point out that Roche-lobe mass transfer and stellar mergers, which are crucial in determining the outcome of binary interaction, and the subsequent evolution 
of the merger are treated by \nbdvi using a simplified treatment of the complex hydrodynamics of stellar mergers, and this may influence individual results. 
Nevertheless, the main conclusions should reflect the real behaviour.
A more detailed discussion of the adopted simplifications and their possible influence can be found in \refs{ssDissBinEvolv}.

\iffigscl
\begin{figure}
\includegraphics[width=\columnwidth]{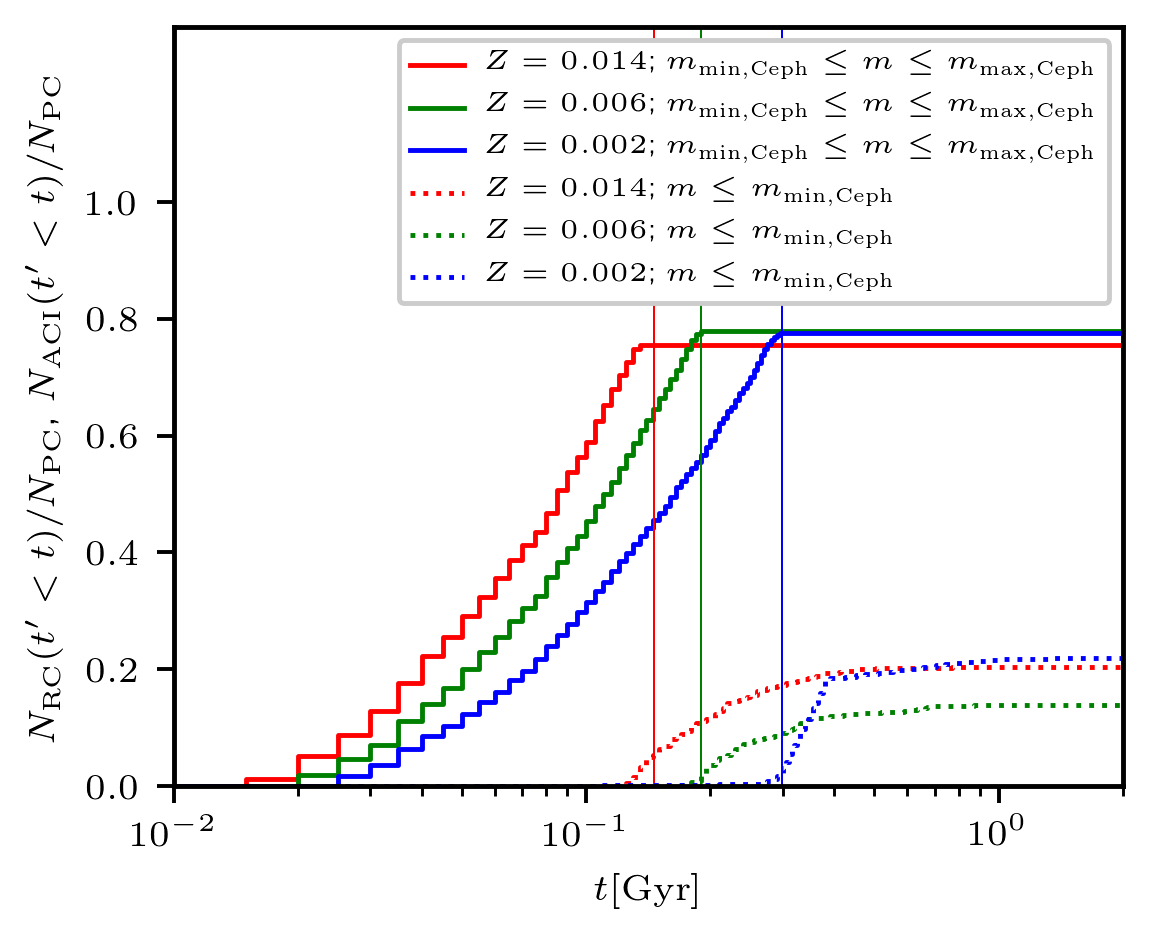}
\caption{
Cumulative number of stars which have become Cepheids 
by the time, $t$, for the control simulations normalized to the number of ProCeps which is taken at the beginning. 
Cepheids originating from ProCeps (solid lines) are plotted separately from the ones originating from stars of $m < m_{\rm min,Ceph}$ (dotted lines). 
The metallicity of the models is indicated by colour. 
The thin vertical solid lines show the time $t_{\rm He,end}$, namely, the time of the least massive non-interacting Cepheid which can occur at a given metallicity; 
all the Cepheids occurring later than $t_{\rm He,end}$ gained some of their mass in interaction with their companion. 
}
\label{fUpCeps}
\end{figure} \else \fi


Next, we studied the Cepheids which originate from stars of the initial mass below the lower mass threshold $m_{\rm min,Ceph}$ of Cepheids (AdCeps of type I), 
but which later increase their mass above the Cepheid mass threshold by a merger or mass transfer event with their companion.
These stars are searched for in the data in the same way as RanCeps; an example of such an evolutionary track is
depicted in the right panel of fig. A.1 in Paper I.
Our models with cluster dynamics terminate at $300 \Myr$, but by this time, they form a comparable number of AdCeps-I as the control models. 
It is probable that the main runs would produce a similar number of AdCeps-I also after $300 \Myr$, so we take the value of 
the relative number of AdCeps-I, $p_{\rm ACI} \equiv N_{\rm ACI}(t' < 2 \Gyr)/N_{\rm PC} (0)$ formed 
in the control runs as a proxy for what is expected to be found observationally. 
The number $N_{\rm ACI}(t' < t)$ of AdCeps-I formed in the control runs by time $t$ is shown by the dotted lines in \reff{fUpCeps}. 
AdCeps-I start appearing slightly before the time, $t_{\rm He, end}$, and 80\% of all AdCeps-I occur before $270 \Myr$, $340 \Myr$, and $390 \Myr$ for $Z = 0.014$, 
$Z = 0.006,$ and $Z = 0.002$, respectively. 
A small number of AdCeps-I occur as late as $1 \Gyr$.  
In contrast, RanCeps occur earlier with no RanCeps occurring after $t_{\rm He, end}$ (solid lines).
The total fractions of AdCeps-I are $p_{\rm ACI} = 0.17$, $0.11,$ and $0.14$ for $Z = 0.014$, $Z = 0.006,$ and $Z = 0.002$, respectively (\reft{ttestRun}).
This means that a stellar population producing $N_{\rm RC}$ Cepheids from ProCeps also produces $p_{\rm ACI} N_{\rm RC}/(1 - p_{\rm EC})$ Cepheids originating 
from stars of mass below $m_{\rm min,Ceph}$. 
Thus, according to the adopted binary evolution recipes, 
approximately one in five observed Cepheids originates from a star with initial mass below the lower 
mass threshold $m_{\rm min,Ceph}$ assumed for Cepheids, and it becomes a Cepheid only due to the mass gain from its companion in a binary.



A small fraction of Cepheids $\lesssim 0.3$\% in our models originate from stars of mass above $m_{\rm max,Ceph}$ (AdCeps type II). 
These objects typically lost some mass to the secondary, but were not prevented by the secondary from becoming Cepheids. 
However, we caution that such rare and complex objects would require a specific in-depth study before firmer conclusions can be drawn.

\subsection{Role of the initial orbital period distribution}

\label{ssWidePer}

We estimated the uncertainty in the initial period distribution by 
comparing the control model of $Z = 0.014$, which has the orbital period distribution of Sana \eqp{eSana}, with the control model for the same metallicity 
but which populates more the longer orbital periods \eqp{eWide}. 
The latter model likely presents an upper limit on the width of the orbital period distribution for ProCeps (cf. \reff{fCephBin}).

ProCeps with the wider distribution of orbital periods \eqp{eWide} experience fewer stellar mergers with $p_{\rm C,coll} = 0.27$ 
($p_{\rm C,coll} = 0.40$ for the control models with Sana's orbital period distribution of \eq{eSana}; 
hereafter in this paragraph, we list the results of the original Sana distribution in parentheses), 
and the fraction of ExCeps is also lower $p_{\rm EC} = 0.17$ ($p_{\rm EC} = 0.24$). 
The less extensive merger history then decreases the fraction of Cepheids of mismatched age to $p_{\rm mma} = 0.14$ ($p_{\rm mma} = 0.20$), and 
it increases the fraction of binary Cepheids to $p_{\rm BC} = 0.59$ ($p_{\rm BC} = 0.50$). 
We do not detect much difference in $p_{\rm cmp}$ nor in $p_{\rm C-C}$. 
The results are summarised in \reft{ttestRun} (last row). 
In total, even though the models with the wider orbital period distribution were taken as the likely upper limit on how wide the 
distribution can be, they do not change the results dramatically; the wider distribution changes any of the quantity listed 
in \reft{ttestRun} by at most $50$\%, but it is generally less for most of the quantities.

\section{On the birth companion frequency among B stars}

\label{sbirthMultiple}

\subsection{Comparison of present models to observations}

\label{ssBinComp}

The estimates of the fraction of Cepheids in triples and quadruples are likely to be lower bounds because our simulations start with subsystems of a maximum of two bodies 
(there are no primordial triples or higher order hierarchies in our models), 
while many young stars of early spectral types are found in more numerous subsystems, for instance, as in the Trapezium in the Orion Nebula Cluster \citep{Preibisch1999}. 
The dynamical encounters in clusters mainly disrupt some of the binaries, but they can also form other binaries, triples and higher order subsystems. 
We recall that our models produce Cepheids with a binary fraction of $42$\% for cluster and field Cepheids combined. 
However, based on \textit{Gaia} DR2 data, \citet{Kervella2019a} provides a lower estimate on Cepheids in binaries as high as $80$\%, 
which implies that assuming even $100$\% of primordial binarity among ProCeps is insufficient to account for the observed binarity of Cepheids.
Possible solutions are primordial triples or higher hierarchies of ProCeps or perhaps a different initial orbital period distribution of ProCeps 
with most ProCeps having orbital periods between $10^2$ to $10^4$ days (avoiding very short orbits to prevent mergers which produce single Cepheids). 
Such an orbital distribution would be in tension with observations of \citet{Sana2012}, suggesting that the more likely explanation 
is mid-B star formation in hierarchical systems.


The fraction of Cepheids in triples is also lower in our models ($\approx 18$\% for Cepheids in clusters and field combined) 
than what is observed ($\approx 35$\% according to \citealt{Evans2005} and \citealt{Kervella2019a}). 
When comparing the companion frequency of Cepheids, which we calculate as $f_{\rm comp} = p_{\rm BC} + p_{\rm tr} + p_{\rm quad}$ 
\footnote{This definition is close to the usual formula 
\begin{equation}
f_{\rm comp} = \frac{B + 2T + 3Q}{S + B + T + Q},
\label{eCSPDef}
\end{equation}
where S, B, T and Q is the number of single stars, binaries, triples and quadruples within the population (\citealt{Duchene2013,Moe2017}, 
see also \citealt{Reipurth1993} for different measures of statistics of multiple stars). 
}
, we obtain $f_{\rm comp} = 0.64$ for the present models, which is almost two times lower than the observed value of $\approx 1.15$ \citep{Evans2005,Kervella2019a}.
The observed value is still a lower estimate because it misses some binaries and triples, and it neglects quadruples. 

Recently, \citet{Evans2020} find that all the Cepheids with a companion closer than $\approx 2000 \Au$ (\logP{ \approx 7.1}) in their sample 
have another inner companion, so the systems are hierarchical triples. 
This differs from the occurrence of close companions to all Cepheids, where only $\approx 30$\% 
of all Cepheids are in spectroscopic binaries \citep{Evans2015,Shetye2024};
namely, the likelihood of a Cepheid to have a close companion is dependent on the presence of another wider companion. 
Clearly, the origin of these triples can be twofold: either primordial or dynamically formed. 
We searched our models for similar triples, and found that only $\approx 5$\% of Cepheids with a companion of orbital period shorter than \logP{ = 7.1} have 
another inner companion. 
Thus, it is unlikely that these outer companions formed dynamically; they are the result of the star formation process.

The necessity of primordial triples or even higher order stable hierarchical systems to explain the observed companion frequency of 
Cepheids points in the same direction as earlier works, which are available 
only for late-type stars \citep{Kroupa1995b,Zwart2004,Kouwenhoven2007,Moeckel2010,Geller2013}, 
where the observed companion frequency cannot be matched by simulations assuming primordial binaries only even if the birth binary fraction is as high as 100\%. 
Our findings support these results and extends them to more massive primaries, where the necessity of primordial hierarchies is important not 
only for the fraction of triples and more numerous systems in evolved clusters, but also for the fraction of binaries. 

\subsection{Estimate of the true birth companion frequency of B stars}

\label{ssEstBirth}

The cluster dynamics and binary evolution can be viewed as mathematical operators transforming 
the birth companion frequency of stars around the spectral type mid-B to the companion frequency of Cepheids 
\footnote{ 
An analogous operator, but acting only during the early evolution of star clusters ($\lesssim 5 \Myr$) was studied by \citet{Marks2011b}.
}.
Cepheids enable detection of less massive and more distant companions than MS mid-B stars because of their narrower spectral lines \citep{Evans2013,Evans2015}.
In our models, the evolutionary operator transforms the initial mid-B star binary fraction of $1.0$  (i.e. companion frequency of $1$) 
to the Cepheid binary fraction of $0.42$. 
Given the observed binary fraction for Cepheids to be $\gtrsim 0.8$, and assuming that the operator is linear (which it is likely not), we 
arrive at an estimate for the birth companion frequency of mid-B stars of $\gtrsim 2.0$, 
namely, all mid-B stars form at least in stable triples on average (with some proportion of quadruples compensated by binaries). 
This estimate is substantially higher than the observed companion frequency for mid-B MS stars ($1.3 \pm 0.2$) \citep{Moe2017}; see also \citet{Abt1990}. 

Setting aside the companions to Cepheids which elude observation, this is still likely a lower estimate. 
More precisely, the evolutionary transformation is not only from mid-B stars $\mapsto$ Cepheids, 
but from mid-B stars and some lower mass stars $\mapsto$ Cepheids because some of the lower mass binaries gain mass by a merger 
with their companion and become AdCeps-I (\refs{ssMergers}). 
The lower mass stars are of late-B and early-A type. 
Approximately $20$ \% of all Cepheids originate from lower mass stars (AdCeps-I),
whereas the remaining $80$ \% of Cepheids originate from the expected mass range (RanCeps, \refs{ssMergers}). 
At the ZAMS, the companion frequency of these stars is then given by the weighted mean of both populations, 
\begin{equation}
f_{\rm comp} = w_1 f_{\rm comp, 1} + w_2 f_{\rm comp, 2},
\label{ebirthFraction}
\end{equation}
where $w_1$ and $f_{\rm comp, 1}$ are, respectively, the number fraction and birth companion frequency for lower mass stars, and 
$w_2$ and $f_{\rm comp, 2}$  are, respectively,  the number fraction and birth companion frequency for ProCeps.
Following the results of \refs{ssMergers}, we set $w_1 = 0.2$ and $w_2 = 0.8$.
The median mass at the ZAMS is $4 \Msun$, and $6.2 \Msun$ for the future AdCeps-I and RanCeps, respectively. 
From \citet[][the right panel of their Fig. 1; see also \citealt{Shatsky2002,Rizzuto2013,Moe2017,Moe2021} for the observational 
basis]{Offner2022}, we extrapolate the observed companion frequencies to be $1.29$ and $1.55$ for the $4 \Msun$, and $6.2 \Msun$ star, respectively. 
Assuming that the true birth ratio of $f_{\rm comp, 2}/f_{\rm comp, 1}$ is the same as the observed one (i.e. $\approx 1.2$), 
\eq{ebirthFraction} provides $f_{\rm comp, 2} = 2.05$ and $f_{\rm comp, 1} = 1.7$ for mid-B and late-B stars, respectively. 
This would indicate a substantially higher companion frequency for late-B stars than currently assumed; a typical late-B star would form as a triple.
On the other hand, assuming that the true birth companion frequency for late-B stars is the same as the observed one (i.e. $f_{\rm comp, 1} = 1.29$), 
\eq{ebirthFraction} provides $f_{\rm comp, 2} = 2.2$, which is even more than the observational estimate for O stars ($f_{\rm comp} = 2.1$; \citealt{Sana2012}). 
In any case, the correction with respect to AdCeps-I increases the estimated companion frequency for mid-B stars above $2$.

Another piece of evidence that the birth companion frequency of mid-B stars is substantially larger than $1.3$ stems from the comparison 
with the observed companion frequency of Cepheids, which is at least $1.15$ (\refs{ssBinComp}). 
Given that $\approx 25$\% of mid-B primaries do not evolve to Cepheids because of the interaction with their companion \citep{Moe2017}, 
the mid-B star companion frequency would decrease from $1.3$ to $1.0$ due to stellar evolution alone, 
i.e. below the value of $1.15$ observed for Cepheids, which is itself a lower estimate. 

The estimate of the birth companion frequency for mid-B stars must be taken with great caution. 
The main limitation is the assumption that the evolutionary operator is linear. 
Including stable hierarchical triples would open more possibilities for interactions between these systems, different interaction cross-sections, 
and also the long-term stability of perturbed hierarchical systems. 
New sets of simulations starting with primordial triples will have to be calculated to constrain the birth companion frequency for B stars more accurately. 

The additional uncertainty is the onset of gas expulsion, $t_{\rm d}$, which we take as $0.6 \Myr$. 
If gas expulsion starts later than this, clusters evolve for longer in their compact sizes, which means that soft binaries 
will be exposed to the dense cluster environment for longer, and thus more of them will be ionised. 
Thus, if $t_{\rm d} > 0.6 \Myr$ ($t_{\rm d} < 0.6 \Myr$), a higher (lower) birth companion frequency will be 
needed to evolve to the same companion frequency of Cepheids than what was obtained for $t_{\rm d} = 0.6 \Myr$. 
The properties of multiple systems at birth that we can deduce from the observations thus intimately depend on the feedback 
physics and compact state of their birth embedded clusters as emphasised by \citet{Kroupa1995a,Kroupa1995b}.

In the estimate of the birth companion frequency, we assume that all triples are intrinsically stable. 
This assumption does not consider primordial unstable triples \citep{Goodwin2005}, which typically decay to a hard binary and a single star. 
Including unstable triples (and possibly also unstable hierarchies of higher order), would raise the mid-B star birth companion frequency above $2$. 


\section{Discussion}

\label{sDiscussion}

\subsection{Stellar evolution of Cepheids in N-body models}

\label{ssDissBinEvolv}

In this paper, we have relied on a simplified picture of the evolution of stars into the evolutionary phase where they may become observable as classical Cepheids. 
The broad strokes of this picture are well understood. 
However, several open questions remain, notably concerning the mass-luminosity (M-L) relation and its dependence on mixing processes, mass-loss, among others. 
N-body simulations integrate stellar evolution via fast lookup functions \citep{Aarseth1999,Hurley2000} since full stellar evolution calculations, 
not to mention detailed modelling of binary evolution, would be prohibitively expensive. 
The fast lookup functions \citep{Hurley2002} in \nbdvi depend on a set of detailed 1D stellar model calculations and notably incorporate 
tides that ensures realistic orbital evolution of close-in binaries on their way to the Cepheid stage. 
However, the detailed modelling of stellar mergers is a very complex subject with many uncertainties and our results, especially for rare cases or the properties of merger products, should be interpreted very carefully. 
For details on binary evolution and the uncertainties involved in modelling the common envelope phase, we refer to \citet{Hurley2002}.
Since predictions will generally differ between stellar evolution codes, a couple of warnings are in order.

Firstly, stellar ages are always model-dependent, whereas the dynamical timescale of cluster evolution only indirectly depends on the stellar evolutionary timescale (e.g. mass transfer triggered by evolution of the primary). 
Since the ages of Cepheids depend strongly on input physics \citep{Bono2005,Anderson2016b,DeSomma2021}, care must be taken when interpreting the results presented here in combination with the output of stellar evolution models, such as period-age relations for Cepheids. 
For example, stellar evolution models featuring rotation predict Cepheids to occur at ages that are approximately 60-100\% older than models that do not include the rejuvenating effects of stellar rotation on the MS \citep{Anderson2016b}. 
Since such changes are relative within an age scale set for a given set of models, this does not affect the statement above 
concerning apparent ages due to mergers or mass transfer events.
However, the dynamical time affecting cluster evolution and the stellar evolution time scale causing evolution-driven interactions could be systematically at odds.

Secondly, the mass range within which Cepheids occur depends on stellar evolution models. 
We here have considered Cepheids within the mass range where the stellar evolution incorporated within \nbdvi predicts blue loops 
to intersect with a theoretically predicted instability strip \citep{Anderson2016b} that has been shown to agree closely with observed Cepheid 
populations \citep{Groenewegen2020,Espinoza2024}.
This ensures internal consistency in our treatment of Cepheids. 
However, this leads to more massive and younger (mixing effects aside, cf. immediately above) Cepheids than those predicted by other models, 
notably ones shown not to be subject to the long-standing Cepheid mass discrepancy \citep[e.g.][]{Caputo2005,Anderson2014}. 
In particular, the minimum mass of Cepheids is correlated with the extent of the blue loops, 
which are particularly uncertain stellar evolutionary effect \citep[e.g.][]{Walmswell2015}. 
However, differences in stellar model masses are of order $10$ to $20$ \% and thus clearly subdominant given the factor of $\approx 3$ in mass covered 
by Cepheids ($\approx 4$ to $12 \Msun$). 
While caution is warranted near the extremes of the mass range, most trends reported here are expected to be rather insensitive to the problem of mass scale.

\subsection{Initial binary population of B stars}

\label{ssInitBinPop}

As we mention in \refs{sInitCond}, we assume that the initial orbital period distribution of B stars is close to the observed one. 
This assumption is adopted mainly because of the lack of previous theoretical works on this topic. 
\refs{ssBin} shows that the properties of binaries evolve substantially due to cluster dynamics, 
with many longer orbital period binaries (\logP{\gtrsim 4}) being softened, which usually leads to disruption (upper row of \reff{fOrbPerChange}). 
If the initial period distribution were more populated at longer orbital periods (\logP{\in (\approx 5, \approx 8)}; analogously to the shape of the initial
period distribution for late type stars as found by \citealt{Kroupa1995b}) than observation by \citet{Sana2012}, most
of the widest binaries would be disrupted early. 
This would produce more single Cepheids, and thus decrease the fraction of Cepheids in binaries, the fraction of Cepheids which 
have merged or those which are of different age than expected from their current mass. 

On the other hand, the estimated initial companion frequency of $2$ (\refs{ssEstBirth}) implies very different initial conditions than estimated here. 
We can view the present paper as the first iteration towards understanding the true birth companion frequency of B stars. 
The second iteration is to perform N-body models of star clusters with mid-B stars initially in triples and quadruples, 
and then to compare the outcome with available observations, and possibly adjust the initial conditions again to perform another iteration until convergence is reached. 

The choice of initial conditions for multiple subsystems is more challenging than for binary stars because 
of the possible correlations between the orbital parameters of the 
inner and outer bodies (see, e.g. \citealt{White2001} for circumstellar disc correlation in early type pre-MS stars).
The presence of initial triples or quadruples likely increases the probability of collisions between stars (e.g. by the Kozai-Lidov mechanism) and interaction 
between the subsystems because the subsystems would mass segregate to the cluster core early, where they would exchange 
some of their members, while ejecting the others as single stars or hard  binaries \citep{Perets2012}.

The lack of primordial triples in our simulations implies that our results regarding the fractions of Cepheids in binaries, triples and quadruples 
underestimate their real values in absolute numbers. 
Despite this, the dependence of these quantities on cluster mass and cluster or field environment (as dealt with in 
Sects. \ref{ssBin} and \ref{ssTriQ}) is instructive and likely follows the real trends.

The present work neglects Cepheids during their first crossing through the instability strip because of the short duration of the event. 
Although the more numerous low mass stars (which do not have second and third instability strip crossing) are expected to cross 
the instability strip once, and spend there a longer time than the stars that evolve from ProCeps, observational evidence suggests that these objects are rare.
Among more than 3600 Milky Way classical Cepheids \citep{Pietrukowicz2021}, there are only four whose rapid period changes 
indicate a first instability strip crossing \citep{Turner2006,Fadeyev2014,Anderson2016a} 
and the pulsation periods of these stars do not imply a particularly low mass. 
Since identifying period changes requires very long temporal baselines (and since first crossings are more easily identified 
due to faster period changes), there is only limited room for a large population of low-mass stars that would only enter the instability strip during a first crossing. 
That said, period changes have only been measured for a couple hundred Milky Way Cepheids \citep{Csornyei2022}, so it is not excluded 
that individual such cases could be found once sufficiently long temporal baselines become available for all known Milky Way Cepheids.
A similar paucity of Cepheids on their first crossing is observed in the Large Magellanic Cloud \citep{RodriguezSegovia2022}, 
which has a lower metallicity ($Z=0.006$) than Milky Way ($Z = 0.014$).
The small population of observed Cepheids on the first crossing implies that they cannot noticeably influence the companion frequency of Cepheids 
and the estimate on the birth multiplicity of B stars.

\subsection{The dependence on galactocentric radius and metallicity}

\label{ssRadZmet}

The dissolution rate of star clusters is a strong function of the cluster orbital radius, $R_{\rm g}$ \citep[e.g.][]{Baumgardt2003}. 
It means that the cluster density and the time interval for which binaries are exposed to cluster environment depend also on $R_{\rm g}$. 
However, we do not detect any clear trend on galactic radius among the properties of binaries, triples or mergers 
studied in this work (see \reft{ttestRun}). 
This result is in contrast to the radial dependence of the fraction of all Cepheids in clusters, 
which increases by a factor of $1.2$ as $R_{\rm g}$ increases from $4 \Kpc$ to $12 \Kpc$ (Paper I). 
The probable explanation for the independence of binary properties of $R_{\rm g}$ 
is that most of the binary dynamical evolution (i.e. due to stellar encounters in the cluster environment)   
occurs during the earliest stage of cluster evolution (before gas expulsion at $t_{\rm d} \approx 0.6 \Myr$ or shortly afterwards), 
which is not influenced by the external field of the galaxy. 
The explanation is in line with numerical works \citep[e.g.][]{Marks2012} and observations \citep[e.g.][]{Bouvier1997,Patience1998,Kraus2011,Geller2012,Moe2017}, 
which indicate that the statistical properties of binary stars in clusters stay largely constant from the age of several Myr to the age of several Gyr.

There is a very weak dependence of the binary properties on metallicity (\reft{ttestRun}). 
As the metallicity decreases, Cepheids reach larger radii \citep{Anderson2016a}, which slightly increases the merger rate with 
their companions (higher $p_{\rm C,coll,\{in,out\}}$ in \reft{ttestRun}). 
The lower minimum Cepheid mass, $m_{\rm min,Ceph}$, at lower $Z$ implies that a higher fraction of Cepheids originate as the secondary, 
while the primary had become a compact object when the secondary becomes a Cepheid. 
Thus, the fraction of Cepheids with a compact companion $p_{\rm cmp}$ increases with decreasing metallicity (\reft{ttestRun}). 
We note that the fraction of Cepheids with mismatched age, $p_{\rm mma}$, decreases with decreasing $Z$ substantially 
in control models according to the Table, but this is artificial because the simulations 
terminate at $300 \Myr$, while the mergers that evolve into Cepheids live longer for these metallicities. 
The control models, which terminate at $3 \Gyr$ show no trend of $p_{\rm mma}$ with metallicity (\reft{ttestRun}).

\section{Summary}

\label{sSummary}


We study the properties of stellar companions to Cepheids by means of direct N-body simulations. 
These models start when all stars (including mid-B stars that evolve into Cepheids) 
form in open star clusters as binaries (but not more numerous subsystems) on the ZAMS.
The binaries are influenced not only by binary star evolution (which is modelled according to 
the recipes of \citealt{Hurley2002}), but also by changes of their orbital parameters due to 
interactions with other stars in the cluster, including dynamical exchanges, binary hardenings, and ionisations. 
We calculate the properties of binaries and multiple systems of these stars at the time when they evolve to Cepheids, 
and discuss the implications for the birth companion frequency of mid-B stars. 
The influence of the dynamical cluster environment is explored here for the first time; 
the only other population synthesis of Cepheids, which is due to \citet{Neilson2015a} and \citet{Karczmarek2022} consider only stellar evolution of isolated binaries.
The main results of the modelling are summarised below.

\begin{itemize}

\item
During binary evolution, many ProCeps interact with their companions by coalescence, Roche-lobe mass transfer or common envelope evolution.
Approximately $30$\% of ProCeps are prevented from becoming Cepheids due to binary evolution. 
In contrast, mass gain in lower mass stars enable some of the stars to become Cepheids even 
though they would not have been able to do so had they evolved in isolation. 
Approximately one in five Cepheids actually originates from stars with mass below the threshold mass for isolated single Cepheids. 
Cepheids from these stars appear at a later time (some of them even at the age of $\gtrsim 500 \Myr$) than expected for Cepheids originating from ProCeps. 

\item
A significant fraction of Cepheids ($\approx 45$\% of Cepheids in clusters and $\approx 36$\% of Cepheids in the field) have merged with their companion; 
however, this does not prevent them from becoming Cepheids. 
Most of the mergers ($60$ to $70$\%) occur immediately after the TAMS (\reff{ftcoal}). 
The larger fraction of mergers in clusters is attributed to dynamical hardening. 
As a result of mergers or mass transfer in binaries, $15$\% to $30$\% of all Cepheids in clusters occur at a different time (more different than by $40$\%) 
than expected from their stellar mass and the cluster age (\reff{ftmmaRatio}). 

\item
The percentage of Cepheids in a binary is substantially higher in clusters ($\approx 60$\%) than in the field ($\approx 35$\%).
Cepheids also have different orbital period distribution in different environments as the result of cluster dynamics (\refs{sssChangesPeriods} and \reff{fOrbPerChange}), 
with $\approx 10$\% of all binary Cepheids having exchanged their companion since the time of their formation (\refs{sssDynExch}). 


\item
Approximately $3$ to $5$\% of all Cepheids have a compact companion (\refs{sssCompactComp}). 
These binaries have a lower mass ratio ($q \approx 0.2$) than all Cepheids in binaries ($q \approx 0.5$).

\item
Binaries where both the primary and secondary are Cepheids at the same time are found in $\approx 1.3$\% of all Cepheids (\refs{sssBothCeps}).

\item
Dynamical evolution in clusters substantially increases the eccentricities of binaries which have initially \logP{\gtrsim 3} and $e \lesssim 0.5$, 
forming a desert in the corresponding area of the $\log_{\rm 10}(P)$ - $e$ plane (upper right panel of \reff{fEccPlane}). 
This transforms the eccentricity distribution from decreasing with $e$ (at the ZAMS) to being flat (when the stars become Cepheids; left panel of \reff{fperDist_v3}). 
The mass ratio distribution between the secondary and primary is not significantly influenced by cluster dynamics.

\item
The percentage of Cepheids in dynamically formed triples is larger in the less massive clusters than in the more massive ones (right panel of \reff{fBinHiaCep}). 
In total, $30$\% of Cepheids in clusters and $10$\% of Cepheids in the field are in triples (\refs{ssTriQ}). 
A similar dependence is seen in quadruples with $5$\% and $3$\% incidence among Cepheids in clusters and in the field, respectively.
Some of the triples can trigger coalescence of the inner binary by increasing its eccentricity via the Kozai-Lidov mechanism. 
A movie showing an example of dynamical formation and disruption of a multiple system of four Cepheids is attached in the accompanying online material.

\item
In addition to stellar evolution, cluster dynamics decreases the fraction of all Cepheids (i.e. in clusters and field combined) in binaries 
($p_{\rm BC} = 0.42$ instead of $p_{\rm BC} = 0.50$ in control models; see \reft{ttestRun}), and it  
decreases the fraction of Cepheids with compact companions ($p_{\rm cmp} = 0.03$ instead of $p_{\rm cmp} = 0.05$). 
On the other hand, cluster dynamics causes more Cepheids of mismatched age ($p_{\rm mma} = 0.27$ instead of $p_{\rm mma} = 0.20$), 
and prevents more ProCeps from becoming Cepheids ($p_{\rm EC} = 0.30$ instead of $p_{\rm EC} = 0.24$).


\item
Binary properties are unaffected by the position of the cluster within its galaxy (\refs{ssRadZmet}).
In lower metallicity environment, there are more Cepheids with compact companions and slightly more Cepheids which have merged with their companion, 
but the overall impact of metallicity is minor (\refs{ssRadZmet}).

\item
Comparing the companion frequency in present models ($\approx 0.64$) with the observed value ($\gtrsim 1.15$), 
we suggest that mid-B stars form with companion frequency substantially larger than $1$, and probably larger than $2$ (\refs{ssEstBirth}). 
This implies that mid-B stars typically form in triples and even more numerous subsystems, most of which are stable on a timescale of at least $\approx 1 \Myr$.


\item
The estimated true companion frequency of mid-B stars thus significantly differs from the value assumed in our work (i.e. $f_{\rm comp} = 1$). 
Starting our simulations with $f_{\rm comp} = 2$ would likely result in different properties of binary, triple, and multiple Cepheids than reported here, 
with mergers, Roche-lobe mass transfer events, hardenings, exchanges, and ionizations being even more abundant. 
However, the dependence of binary properties on the cluster mass and field or cluster environment likely reflects real trends.
The present simulations likely provide a lower estimate on these quantities. 

\end{itemize}

\begin{acknowledgements}
We thank the referee for the constructive and very detailed comments, which improved the quality and clarity of the paper.
FD appreciates the ESO Visiting Scientist Programme, which made this project possible via his stay at ESO Garching. 
We would like to thank to Sverre Aarseth for continuously developing the code \nbdvid, 
and to Sambaran Banerjee and Seungkyung Oh for their advice about specifying integration parameters for the simulations. 
FD thanks to Stefanie Walch for her encouragement to work on the project while being employed in her group, and acknowledges support of the 
Deutsche Forschungsgemainschaft (DFG) through SFB 956 "The conditions and impact of star formation" (sub-project C5).
RIA is funded by the SNSF Eccellenza Professorial Fellowship PCEFP2\_194638
and acknowledges funding from the European Research Council (ERC) under the European Union's Horizon 2020 research and innovation programme (Grant Agreement No. 947660).
FD and PK acknowledge support through grant 20-21855S from the Grant Agency of the Czech Republic. 
PK acknowledges support through the DAAD-Easter-Europe Exchange grant at Bonn University and corresponding support from Charles University. 
We appreciate the support of the ESO IT team, which was vital for performing the presented numerical models. 
This research made use of the Matplotlib Python Package \citep{matplotlib2007}.
\end{acknowledgements}

%
%




\bibliographystyle{aa} 
\bibliography{cepheidsEscape} 


\begin{appendix}
 
\newcolumntype{C}[1]{%
 >{\vbox to 8ex\bgroup\vfill\centering\arraybackslash}%
 p{#1}%
 <{\vskip-\baselineskip\vfill\egroup}}
 
\section{Comparison of terminology with Paper I}

\begin{sidewaystable}
\caption{
Translation between the terminology used in Paper I and the present paper (Paper II). 
}
\label{tTerminology}

\begin{tabular}{c|c|C{0.7\textwidth}}
Paper I & Paper II & description \\
\hline
ProCep & ProCep & A star with the ZAMS mass in the interval $(m_{\rm min,Ceph}, m_{\rm max,Ceph})$. \\
\hline
UpCep & AdCep-I & A Cepheid which had the ZAMS mass below $m_{\rm min,Ceph}$, but evolved to a Cepheid because of a mass gain due to a mass transfer event or merger. \\
\hline
CanCep & RanCep & A Cepheid which had the ZAMS mass in the interval $(m_{\rm min,Ceph}, m_{\rm max,Ceph})$ (i.e. it formed as a ProCep).
                  A fraction of these stars have experienced a mass transfer event or merger with its companion, while others 
                  have not. \\
\hline
(\textit{no equivalent}) & CanCep & A Cepheid which had ZAMS mass in the interval $(m_{\rm min,Ceph}, m_{\rm max,Ceph})$ (i.e. it formed as a ProCep), 
  and which has neither experienced a mass transfer event nor merger.
\end{tabular}
\tablefoot{
The terms listed on each line are equivalent; their definition is provided in the right column. 
Note that CanCeps have a different meaning in paper I than in paper II, the other terms are either equivalent (ProCeps) or unique to paper I (UpCeps). 
The Table does not list the terms which are unique to paper II because they cannot be confused with the terms used in paper I. 
See \refs{sChannels} and \reff{ftimeCartoonCeph} for their full list.
}
\end{sidewaystable}

\end{appendix}

\end{document}